\documentclass[useAMS,usenatbib]{mnras}
\usepackage{graphicx}\usepackage{epsfig}\usepackage{epsf}
\usepackage{amsmath}
\usepackage{threeparttable}
\usepackage{amssymb}
\usepackage{stfloats}
\usepackage{soul}
\usepackage[usenames]{color}
\usepackage{tree-dvips}

\definecolor{green}{rgb}{0.2,0.6,0.}
\definecolor{rust}{rgb}{0.7,0.1,0.}

\usepackage{adjustbox}
\usepackage{longtable}
\usepackage{multicol}
\usepackage{xltabular}
\usepackage{lipsum}
\usepackage{orcidlink}
\DeclareRobustCommand{\VAN}[3]{#2} 
\usepackage{hyperref}
\usepackage[caption=false]{subfig} 
\title[LMC LBVs]{Kinematics of Luminous Blue Variables in the Large Magellanic Cloud}

\author[M. Aghakhanloo et al.]{Mojgan Aghakhanloo \orcidlink{0000-0001-8341-3940},$^{1}$\thanks{E-mail:
aghakhanloo@arizona.edu} Nathan Smith \orcidlink{0000-0001-5510-2424},$^1$ Jennifer Andrews \orcidlink{0000-0003-0123-0062},$^{2}$ Knut Olsen,$^3$ \newauthor Gurtina Besla \orcidlink{0000-0003-0715-2173},$^1$ Yumi Choi \orcidlink{0000-0003-1680-1884}$^4$\\ 
$^1$ Steward Observatory, University of Arizona, 933 N. Cherry Ave., Tucson, AZ 85721, USA \\
$^2$ Gemini Observatory, 670 N. Aohoku Place, Hilo, Hawaii, 96720, USA \\  
$^3$ National Optical Astronomy Observatory, 950 N. Cherry Ave., Tucson, AZ 85719, USA \\ $^4$ Space Telescope Science Institute, 3700 San Martin Drive, Baltimore, MD 21218, USA}

\begin{document}

\pagerange{\pageref{firstpage}--\pageref{lastpage}} \pubyear{2022}
\maketitle
\label{firstpage}

\begin{abstract}
    We study the kinematics of luminous blue variables (LBVs) in the Large Magellanic Cloud (LMC). Using high-resolution spectra, we measure the systemic radial velocities for a sample of 16 LBVs and LBV candidates. In order to measure the net motion of LBVs compared to their local environments, we subtract the projected line-of-sight velocity at the same location derived from the rotation curve model of the LMC. Using nebular and wind emission lines,
    we infer a velocity dispersion for LBVs of $40.0^{+9.9}_{-6.6}$ km s$^{-1}$. To put LBVs in context with other evolved massive stars, we compare this to red supergiants (RSGs) in the LMC, which have a significantly smaller velocity dispersion of $16.5^{+0.4}_{-0.6}$ km s$^{-1}$. Moreover, 33\% of LBVs have radial velocities of more than 25 km s$^{-1}$, while only 9\% of RSG have such high velocities. This suggests that LBVs include more runaways than the population of stars that evolves to become RSGs, indicating that LBVs are preferentially kicked by a companion's supernova explosion as compared to other evolved massive stars. Our investigation reveals other interesting clues about LBVs in the LMC as well. We find that radial velocities and widths of {\it emission} lines for each target remain constant over several epochs, whereas measured {\it absorption} lines exhibit highly variable radial velocities for R110, R81, S Dor, Sk-69$^\circ$142a, and Sk-69$^\circ$279. These five LBVs probably have a binary companion.  Additionally, we find that Sk-69$^\circ$142a experienced its second outburst in 2019 September, shifting its status from candidate to confirmed LBV. 
\end{abstract}

\begin{keywords}
galaxies: individual (LMC) – galaxies: stellar content – Magellanic Clouds – stars: evolution – stars: massive.
\end{keywords}

\section{INTRODUCTION}\label{sec:intro}
Luminous blue variables (LBVs) are the brightest blue irregular variable stars in any large star-forming galaxy \citep{HS53,T68}. These were originally known as the “Hubble-Sandage variables” but later \citet{C84} related them to the famous Galactic objects like P Cygni, and $\eta$ Carinae, plus other hot supergiants with strong mass loss, and grouped them together as “LBVs”. LBVs represent an observational phenomenon where the stars experience irregular photometric variability or major eruptive mass loss events \citep{V01,S11,S17}, but the physical cause of LBV eruptions is not yet clear. Stars that resemble LBVs in their measured physical properties and the morphology of their spectra --- but have not yet been caught showing the signature variability of LBVs  --- are called LBV candidates.

In the traditional single-star scenario, stars above $\sim$30 M$_{\odot}$ pass through an LBV phase. In this brief phase, they experience eruptive mass-loss as a means to transition from a hydrogen-rich star to an H-poor Wolf--Rayet (WR) star \citep{C75,C84,hd94}. If LBVs mark
a brief transitional phase for single stars at the end of the main sequence and before
core-He burning WR stars, then they should be found in young clusters near other comparably massive
O-type stars.  However, \citet{st15} found that LBVs are relatively
isolated from O-type stars, and even farther away from O stars than
the WR stars are. Given their isolation, \citet{st15} concluded that LBVs should be products of binary evolution. In a follow-up theoretical study, \citet{mojgan17} developed models for the passive dispersal of clusters that could quantitatively explain the observed separation distribution of O-type stars with a standard drift velocity, but could not account for the locations of LBVs with the same cluster dispersal. LBVs require either a faster dispersal velocity or longer lifetimes than expected. Therefore, the LBV phenomenon is inconsistent with a single-star scenario and is more consistent with binary scenarios where LBVs are massive blue stragglers \citep{st15,mojgan17}.

Despite the recent evidence that observationally and theoretically LBV’s spatial distribution is consistent with the binary evolution, some interpretations were proposed favoring the traditional single-star scenario \citep{H16, Aadland18}. \cite{H16} suggested that separating the less luminous and more luminous LBVs into two samples would make their separations consistent with O-type stars, although they did not also split the O stars in a consistent way. However, \citet{S16} showed that separating LBVs into different subgroups does not change the result that LBVs are isolated from O-type stars. Later, \citet{Aadland18} also proposed that LBVs’ environment is consistent with the single star scenario. However, the sample used in their study contains an older population of evolved bright blue stars, which should not be compared with the young population of LBVs \citep{S19bs}. \citet{K19} also suggested that LBVs in \citet{Aadland18} catalog are misclassified and the sample contains many B[e] supergiants.

Two binary evolutionary scenarios may potentially explain LBVs' isolation from O-type stars: 1) LBVs could be mass gainers in a binary system where the primary mass donor has already exploded as a supernova (SN), imparting the companion with a significant net motion after the explosion. In this case, LBVs may be isolated from O-type stars because they are runaways. 2) They may be rejuvenated either by mass accretion in a binary (without a significant kick from a companion's SN), or by being the product of a stellar merger.  These two options have been discussed in several recent papers \citep{J14, st15, mojgan17,SN11,S16,S19bs,smith2016merger,smith2016sn09ip,smith2018}. The mass gainer or merger process can substantially increase the star's mass and luminosity. As initially lower-mass stars, the core-H burning lifetimes of LBVs may have been much longer than the main-sequence lifetime expected for the more luminous star that they become. Therefore, due to longer ages, LBVs will have enough time to wander far away from O-type stars, or their O-type siblings may already be dead.

In addition to the binary SN scenario, LBVs can, like any massive stars become runaway stars via dynamical ejection as well. In this scenario, stars get kicked out of the cluster via interaction with a massive binary located at the center of the cluster \citep{B61,P67,L88}. 
The dynamical ejection can produce runaway stars with higher velocities in comparison to the binary supernova scenario, up to 200 km s$^{-1}$ \citep{P12} or even up to $\sim$400 km s$^{-1}$ \citep{O16}. \citet{Renzo} found that the majority of the kick velocities by the supernova explosion of the companion stars are relatively slow ($<$ 30 km s$^{-1}$). However, an important point is that dynamical ejections cause most runaways early in the cluster evolution before the first SN explosion \citep{O16}. Therefore, these runaways will also affect the larger O-type star population and other evolved massive stars like red supergiants (RSGs) and WR stars, to which LBVs have been compared. Indeed, this probably explains the tail of large separations in the observed separation distribution of O-type stars \citep{st15}.  But there is no clear reason why dynamical ejections would {\it preferentially} eject most of the LBVs --- it applies to all massive stars early in life, and so this mechanism cannot explain why the median separation of LBVs is so much larger than that of O-type stars.  Dynamical ejections could, however, contribute a few runaways among the larger population of LBVs.

If LBVs arise from binary interaction, their initial masses can be at least $\sim$19 M$_{\odot}$ and their isolation from O stars is consistent with a wide range of kick velocities, anywhere from 0 to $\sim$105 km s$^{-1}$ \citep{mojgan17}. In order to quantitatively account for the observed dispersal of LBVs, there is a competition between kicks and rejuvenation, in the sense that systematically higher kick velocities would require less rejuvenation through mass gainers and mergers to achieve the same observed dispersal. Understanding the kinematics of LBVs is therefore an important key to unraveling the source of their apparent isolation.

There are 25 LBVs in our Galaxy \citep{S19}, 19 in the nearest galaxies, the Large Magellanic Cloud (LMC) and Small Magellanic Cloud \citep[SMC;][]{st15,S16,S19bs}, and 33 in M31 and M33 \citep{H17}. (We are including both confirmed LBVs and LBV candidates.) In this work, we focus on the kinematics of LBVs in the LMC to constrain whether they appear isolated because they are predominantly rejuvenated blue stragglers, or because they have received a significant kick from their companion.  \citet{S16} studied the kinematics of LBVs in M31 and M33, and found that they had higher median velocities as compared to a sample of RSGs that was consistent with moderate kicks.

One way to test whether LBVs are runaways or have been rejuvenated is to study their radial velocity with respect to their local environment. However, estimating the radial velocity of LBVs is tricky because their strong winds and irregular variability that results from their instability may affect the centroids of absorption and emission lines. Orbital motion in a binary also may affect their radial velocity, because a majority (70-100\%) of massive stars are in binary systems \citep{S12,M17}.

The main goal of this paper is to estimate systemic radial velocities of LBVs in order to constrain their net motion with respect to their surroundings.
First, we describe new spectral observations and the data reduction process, and then the measurement of radial velocities from spectra in section~\ref{sec:obs}. In section \ref{sec:dispersions}, we examine the resulting velocity dispersion of LBVs and RSGs after correcting for the rotation of the LMC, and we use a Bayesian inference technique to determine the velocity dispersion of LBVs as well as RSGs.
We also discuss the implications of the inferred dispersion velocities. Section~\ref{sec:consclusion} presents a summary and direction for further investigation. 

\section{OBSERVATIONS}\label{sec:obs}
\subsection{MIKE Echelle Spectra}
We obtained high-resolution spectra of LBVs in the LMC using the Magellan Inamori Kyocera Echelle (MIKE) spectrograph, mounted on the 6.5m Clay Telescope of the Magellan observatory at Las Campanas, Chile.  MIKE is a double echelle spectrograph designed for use at the Magellan Telescopes \citep{B03}.  We obtained new MIKE spectra of 9 confirmed LBVs and 7 LBV candidates in the LMC, including two targets for which no radial velocity had previously been reported in the literature. The LBV sample used in this work is adopted from \cite{st15}. As we discussed in section~\ref{sec:intro}, \cite{st15} revealed that LMC LBVs are
isolated from O-type stars.  Therefore, in this work, we obtained MIKE spectra of the same sample of LBVs to determine whether these LBVs are isolated from O stars because of their peculiar kinematics. Table~\ref{tab:tab1} lists the observation dates for each target. 

MIKE provides high-resolution cross-dispersed echelle spectra from $\sim$ 3200 \AA \ to $\sim$ 10000 \AA \ in two channels. We  used the 1 arcsec $\times$ 5 arcsec slit for all stars, yielding a spectral resolution R = $\lambda$/$\Delta\lambda$ of about 30,000 or roughly 10 km s$^{-1}$ in the blue arm, and roughly R = 20,000 or 15 km s$^{-1}$ in the red arm. The spectra were reduced using the latest version of the MIKE pipeline written by D. Kelson\footnote{\url{http://code.obs.carnegiescience.edu/mike/}}. Further details on the heliocentric-corrected radial velocity measurements are described in the next section.

{\it Gaia} Early Data Release 3 \citep[EDR3;][]{g20} also provides proper motion measurements for sources in the LMC. However, most of the LMC LBVs have either negative parallaxes or large uncertainties due to crowding, high extinction, and binarity \citep{L18}. Therefore, we do not include {\it Gaia} EDR3 proper motions in this work.

\subsection{Measuring Radial Velocities}\label{sec:velocities}
Table~\ref{tab:tab1} presents the heliocentric radial velocities of the LBVs measured using our high-resolution echelle spectra. We measure the radial velocities of LBVs in the LMC using IRAF\footnote{IRAF is distributed by the National Optical Astronomy Observatory, which is operated by the Association of Universities for Research in Astronomy (AURA) under a cooperative agreement with the National Science Foundation.} by fitting Gaussian profiles for individually selected absorption or emission lines. To estimate the radial velocities, we adopt the rest air wavelengths from the NIST Atomic Spectra Database \citep{K16} and Peter van Hoof's Atomic Line List\footnote{ \url{http://www.pa.uky.edu/~peter/atomic/}}.

Our goal is to measure the average systemic velocity of each LBV or binary system containing an LBV in order to compare this with the expected velocity for its location in the LMC.  However, measuring accurate systemic radial velocities for LBVs is complicated because they are unstable stars with irregular variability, where different lines form at different radii in dense winds that may have high optical depths and may be changing with time.  For some lines, emission and absorption velocities are skewed by high optical depths in the wind, as is obviously the case for lines with a P Cygni profile.  Moreover, if an LBV is in a binary system, the observed radial velocity might be altered by Doppler shifts introduced by orbital motion.  To explore and mitigate these effects, we chose three different types of measurements to make: (1) absorption lines formed near the photosphere, (2) emission lines in the wind, and (3) average velocities for forbidden lines formed in a circumstellar nebula (if present).   

\begin{figure}
\includegraphics[width=\linewidth]{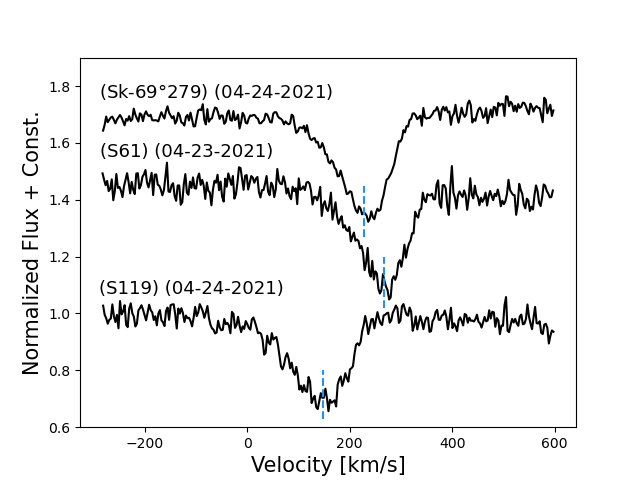}
\caption{Example of the Si~{\sc IV} $\lambda$4089 \AA \ absorption line used to estimate the radial velocity of several LBVs. Heliocentric radial velocities are marked in dashed blue lines.}\label{fig:HeIGaussianFit}
\end{figure}

{\it Absorption lines:} We measure absorption lines using transitions that are expected to form in the photosphere or deep in the wind near the photosphere, which exhibit centroids consistent with zero velocity in radiative transfer simulations of LBV spectra (J.\ Groh, private communication; see also \citealt{G09,G11}).  Examples of such lines are He~{\sc i}, He~{\sc ii}, N~{\sc iii}, Ca~{\sc ii}, Si~{\sc ii}, Si~{\sc iii}, Si~{\sc iv}, and Mg~{\sc ii}, with different lines being suitable depending on the effective temperature ($T_{\rm eff}$) of each star. Fig.~\ref{fig:HeIGaussianFit} shows examples of Si~{\sc iv} $\lambda$4089 \AA \ absorption for a few LBVs.  Obviously, the degree to which this absorption deviates from the true systemic velocity may vary from object to object depending on the density and speed of the wind.  We note that there is a consistent trend that absorption lines are more blueshifted than emission lines (discussed next), suggesting that absorption features are significantly influenced by wind absorption, so we do not favor these absorption lines as indicators of the systemic velocity.

\begin{figure}
\includegraphics[width=\linewidth]{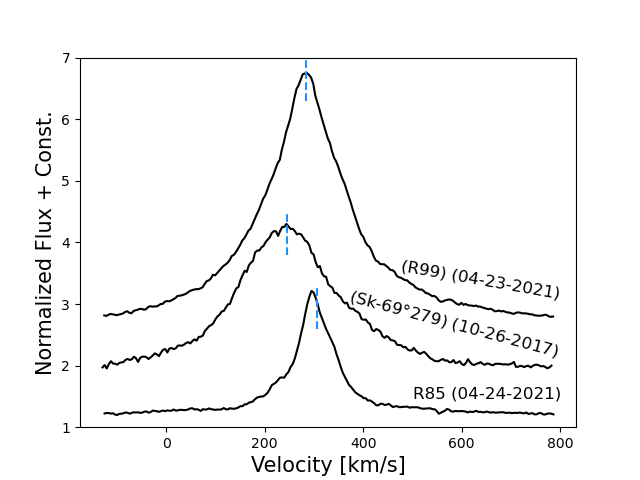}
\caption{Example of the H$\alpha$ $\lambda$6563 \AA \ emission line used to estimate the radial velocity of LBVs. Heliocentric radial velocities are marked in dashed blue lines.}\label{fig:Halpha}
\end{figure}

\begin{figure}
\includegraphics[width=\linewidth]{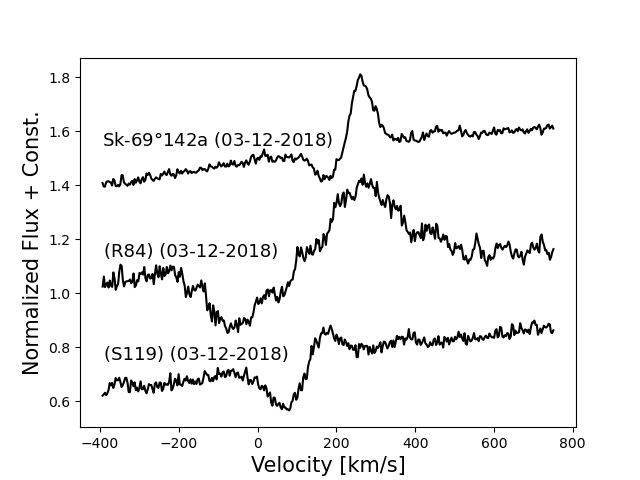}
\caption{Examples of the types of spectral lines (He~{\sc i}  $\lambda$7065 \AA) that we avoid in this study because they exhibit P Cygni features, indicating that velocities are strongly impacted by blueshifted wind absorption. Even so, it is clear that S119 has a line centroid that is strongly blueshifted.} \label{fig:Pcygni}
\end{figure}

{\it Wind emission lines:}  Centroids of emission lines formed in the wind, especially lines formed relatively far out in the wind, should provide a good measure of the systemic velocity if the wind is spherical or axisymmetric.  Wind emission might even avoid Doppler shifts from binary orbital motion if the line is formed far enough out in the wind so that it is formed beyond the orbital separation, or in eccentric binary systems where the two stars spend most of their time moving slowly at apastron.  We measure emission-line centroids for several lines including H$\alpha$, Fe~{\sc ii}, [Fe~{\sc ii}], and various other lines.  Indeed, these emission lines produced more consistent radial velocities than absorption lines. Fig.~\ref{fig:Halpha} shows examples of H$\alpha$ emission lines without P Cygni absorption used to estimate the radial velocity of several LBVs, and Fig.~\ref{fig:Pcygni} shows examples of spectra with P Cygni profiles that we avoid using for measuring radial velocities. Appendix \ref{sec:appendix} shows all the emission lines that are used to measure the radial velocity of LBVs.

\onecolumn
\LTcapwidth=\textwidth
\begin{longtable}{cccc}
\caption{List of the LBVs and candidate LBVs (names in parentheses) in the LMC that we  observed using the MIKE echelle spectrograph on the Clay Telescope, the date of observations, the line IDs for features used to measure  velocities, and the radial velocities of the associated absorption or emission lines.} \label{tab:long} \\

\hline \multicolumn{1}{c}{\textbf{Name}} & \multicolumn{1}{c}{\textbf{Obs. Date}} &
\multicolumn{1}{c}{\textbf{Line ID}} & \multicolumn{1}{c}{\textbf{Radial Velocity [km s$^{-1}$]}}  \\ \hline \hline 
\endfirsthead

\multicolumn{4}{c}%
{{\bfseries \tablename\ \thetable{} -- continued from previous page}} \\
\hline \multicolumn{1}{c}{\textbf{Name}} & \multicolumn{1}{c}{\textbf{Obs. Date}} & \multicolumn{1}{c}{\textbf{Line ID}} & \multicolumn{1}{c}{\textbf{Radial Velocity [km s$^{-1}$]}} \\ \hline\hline
\endhead

\hline \multicolumn{4}{r}{{Continued on next page}} \\ \hline
\endfoot

\hline 
\endlastfoot

 S Dor &03-11-2018 &Fe~{\sc ii}, [Fe~{\sc ii}]&  305.1$\pm$0.4$^{eP}$ \\
  & &Si~{\sc ii}&  248.5$\pm$0.6$^{ac}$\\

  & 04-24-2021& Fe~{\sc ii}, [Fe~{\sc ii}]&  305.7$\pm$0.4$^e$\\
   & & Multiple Si~{\sc ii} &  296.8$\pm$0.5$^{ac}$\\
   
  & 08-01-2021& Fe~{\sc ii}, [Fe~{\sc ii}] &  308.7$\pm$0.2$^e$ \\
   & & Multiple Si~{\sc ii} &  301.5$\pm$0.2$^{ac}$\\
   
    & 10-02-2021&   Fe~{\sc ii}, [Fe~{\sc ii}]&  307.5$\pm$0.3$^{ei}$\\
   & &  Si~{\sc ii} &  296.9$\pm$0.3$^a$  \\
   
   & 10-12-2021&  Fe~{\sc ii}, [Fe~{\sc ii}]&  307.1$\pm$0.2$^{ei}$\\
    & &  Si~{\sc ii}&  296.0$\pm$0.1$^a$  \\
   
\hline
R81 &03-11-2018& Multiple Si~{\sc ii} &  255.1$\pm$6.9$^e$ \\
 & & Multiple He~{\sc i} &  298.3$\pm$1.6$^a$ \\

 & 04-24-2021& Multiple Si~{\sc ii} &  253.5$\pm$6.7$^e$ \\
 & & Multiple He~{\sc i} &  229.3$\pm$1.5$^a$ \\
 
 & 08-01-2021& Multiple Si~{\sc ii} &  256.1$\pm$6.3$^e$ \\
 & & Multiple He~{\sc i} &  233.6$\pm$0.3$^a$ \\
 
   & 10-02-2021&  Multiple Si~{\sc ii}&  255.3$\pm$5.6$^e$     \\
   & &  Multiple He~{\sc i} &  237.0$\pm$0.4$^a$ \\
   
   & 10-12-2021&  Multiple Si~{\sc ii}&  253.1$\pm$7.7$^e$   \\
   & &  Multiple He~{\sc i} &  243.1$\pm$0.7$^a$\\
\hline
R110 & 10-25-2017& Fe~{\sc ii} &  267.1$\pm$1.3$^e$\\
 & & Multiple Si~{\sc ii}&  259.4$\pm$0.7$^{ai}$\\

 & 04-25-2021& Fe~{\sc ii}, [N~{\sc ii}] &  272.4$\pm$3.7$^e$\\
 & & Multiple Si~{\sc ii}&  278.1$\pm$0.6$^{ai}$\\
 
 & 08-01-2021& Fe~{\sc ii}, [N~{\sc ii}] &  277.1$\pm$4.1$^e$\\
 & & Multiple Si~{\sc ii}&  281.9$\pm$0.6$^{ai}$\\
 
    & 10-02-2021&   Fe~{\sc ii}, [N~{\sc ii}] &  273.8$\pm$3.9$^e$  \\
   & & Multiple Si~{\sc ii}&  263.3$\pm$0.5$^{ai}$\\
   
    & 10-12-2021&  Fe~{\sc ii}, [N~{\sc ii}] &  274.5$\pm$3.8$^e$  \\
   & & Multiple Si~{\sc ii}&  264.8$\pm$0.5$^{ai}$\\
\hline
R71 & 10-25-2017& [Fe~{\sc ii}] &  193.8$\pm$0.1$^{e}$ \\
 & & Multiple Ca~{\sc ii} &  163.3$\pm$0.1$^{ac}$   \\

&03-11-2018& [Fe~{\sc ii}] &  196.5$\pm$0.1$^{e}$ \\
& & Multiple Ca~{\sc ii} &  163.5$\pm$0.1$^{ac}$ \\
 
 & 04-24-2021& [Fe~{\sc ii}] &  195.2$\pm$0.2$^{e}$ \\
& & Multiple Ca~{\sc ii} &  166.1$\pm$0.1$^{ac}$ \\

& 08-01-2021& [Fe~{\sc ii}] &  197.7$\pm$0.2$^{e}$ \\
& & Multiple Ca~{\sc ii} &  168.8$\pm$0.1$^{ac}$ \\
\hline
MWC112 & 03-12-2018 & Mg~{\sc ii}, He~{\sc ii} &  266.7$\pm$0.1$^{ac}$  \\
 & 04-25-2021 & Mg~{\sc ii},  He~{\sc ii} &  263.3$\pm$0.1$^{ac}$  \\
     & 10-02-2021&  Mg~{\sc ii},  He~{\sc ii} &  268.6$\pm$0.1$^{ac}$  \\
\hline
R85 & 03-12-2018& H$\alpha$ &  304.0$\pm$0.1$^{ec}$\\
 & & Multiple Si~{\sc ii} &  280.9$\pm$0.8$^a$    \\
 
 & 04-24-2021& H$\alpha$ &  305.3$\pm$0.1$^{e}$  \\
 & & Multiple Si~{\sc ii} &  284.1$\pm$0.8$^a$    \\
   
    & 10-02-2021& H$\alpha$ &  306.1$\pm$0.1$^e$   \\
   & &  Multiple Si~{\sc ii} &  279.7$\pm$0.7$^a$  \\
 \hline
R143 & 04-23-2021& Multiple Mg~{\sc ii} &  276.2$\pm$1.1$^{ei}$  \\
 & & Multiple Si~{\sc ii} &  282.9$\pm$1.4$^a$    \\
 & 04-25-2021& Multiple Mg~{\sc ii} &  276.3$\pm$1.0$^{ei}$  \\
 & & Multiple Si~{\sc ii} &  283.9$\pm$0.9$^a$    \\
  & 08-01-2021& Multiple Mg~{\sc ii} &  274.6$\pm$0.4$^{ei}$ \\
 & & Multiple Si~{\sc ii} &  280.6$\pm$0.6$^a$    \\
 \hline
Sk-69$^\circ$142a & 03-12-2018& Si~{\sc ii} &  264.1$\pm$1.2$^e$  \\
& & Multiple Si~{\sc iii} &  225.5$\pm$2.7$^{ap}$ \\

& 04-25-2021& Si~{\sc ii} &  260.1$\pm$1.6$^e$  \\
& & Multiple Si~{\sc iii} &  197.4$\pm$2.9$^{ap}$\\

& 08-01-2021&  Si~{\sc ii} &  259.8$\pm$1.5$^e$ \\
& & Multiple Si~{\sc iii}  &  188.4$\pm$2.3$^{aP}$ \\

    & 10-02-2021&  Si~{\sc ii} &  262.1$\pm$1.1$^e$   \\
   & &  Multiple Si~{\sc iii} &  188.5$\pm$1.9$^{aP}$  \\
   
    & 10-12-2021&  Si~{\sc ii} &  261.3$\pm$1.2$^e$  \\
   & & Multiple Si~{\sc iii} &  187.7$\pm$2.5$^{aP}$ \\

\hline
(R84) & 03-12-2018& H$\alpha$, H$\beta$ & 249.3$\pm$0.2$^e$ \\
 & & Multiple Ca~{\sc ii} & 246.7$\pm$1.0$^a$ \\
 
 & 04-24-2021& H$\alpha$, H$\beta$ & 248.9$\pm$0.3$^e$ \\
 & & Multiple Ca~{\sc ii} & 246.0$\pm$0.8$^a$ \\
   
    & 10-02-2021&  H$\alpha$, H$\beta$ & 253.3$\pm$0.2$^e$    \\
   & &  Multiple Ca~{\sc ii} & 246.3$\pm$0.6$^a$  \\
\hline
(R126) & 03-12-2018& He~{\sc i}, N~{\sc iii} &  258.3$\pm$0.7$^e$ \\
 & 04-25-2021& He~{\sc i}, N~{\sc iii} &  256.3$\pm$1.0$^e$ \\
     & 10-02-2021&  He~{\sc i}, N~{\sc iii} &  258.4$\pm$0.7$^e$   \\
   
\hline
(S119) & 03-12-2018& He~{\sc ii} & 144.0$\pm$0.5$^e$  \\
 & & Multiple Si~{\sc iv}&  147.5$\pm$1.3$^a$  \\
 
 & 04-24-2021&  He~{\sc ii} &  144.6$\pm$1.3$^e$  \\
 & & Multiple Si~{\sc iv}&  147.4$\pm$2.2$^a$  \\
 
  & 08-01-2021&  He~{\sc ii} &  143.4$\pm$0.6$^e$  \\
 & & Multiple Si~{\sc iv}&  149.5$\pm$1.7$^a$  \\
\hline
(Sk-69$^\circ$271) & 10-26-2017& H$\alpha$&  299.3$\pm$1.1$^{ec}$ \\
 & & Multiple He~{\sc i}&  260.6$\pm$1.5$^a$   \\

 & 04-25-2021& H$\alpha$&  301.0$\pm$0.7$^{ec}$   \\
 & & Multiple He~{\sc i}&  258.8$\pm$1.6$^a$   \\
 
   & 10-02-2021&  H$\alpha$&  299.7$\pm$0.5$^{ec}$ \\
     & &  Multiple He~{\sc i}&  262.1$\pm$1.7$^a$  \\
 \hline
(R99)& 04-23-2021& He~{\sc i}, H$\alpha$, H$\beta$, H$\gamma$ &  281.8$\pm$0.7$^e$ \\
& 04-25-2021& He~{\sc i}, H$\alpha$, H$\beta$, H$\gamma$ &  280.7$\pm$0.8$^e$ \\
&  08-01-2021& He~{\sc i}, H$\alpha$, H$\beta$, H$\gamma$ &  279.6$\pm$0.4$^e$ \\
 \hline
(S61)& 04-23-2021& Si~{\sc iv} &  285.1$\pm$2.1$^e$ \\
 & & Si~{\sc iv}, He~{\sc ii} &267.9$\pm$5.6$^a$ \\

& 04-25-2021& Si~{\sc iv} &  284.5$\pm$2.3$^e$ \\
& & He~{\sc ii} &  267.1$\pm$5.1$^a$ \\

&  08-01-2021& Si~{\sc iv} &  283.8$\pm$1.9$^e$ \\
& & Si~{\sc iv}, He~{\sc ii} &  267.8$\pm$5.0$^a$ \\
\hline
(Sk-69$^\circ$279)& 10-26-2017& H$\alpha$ &  246.1$\pm$0.2$^e$ \\
 & & Multiple Si~{\sc iv} &210.3$\pm$2.7$^a$ \\

& 04-24-2021& H$\alpha$ &  254.8$\pm$0.1$^e$ \\
& & Multiple Si~{\sc iv}&  227.7$\pm$1.9$^a$ \\

&  08-01-2021& H$\alpha$ &  252.8$\pm$0.1$^e$\\
& & Multiple Si~{\sc iv}&  224.7$\pm$1.3$^a$ \\

    & 10-02-2021&  H$\alpha$ &  245.6$\pm$0.1$^e$   \\
   & &  Multiple Si~{\sc iv}&  233.0$\pm$1.0$^a$ \\
   
    & 10-12-2021&  H$\alpha$ &  254.7$\pm$0.1$^e$\   \\
   & &  Multiple Si~{\sc iv}&  226.1$\pm$1.6$^a$   \\
\end{longtable}\label{tab:tab1}
\footnotesize{$^e$ Emission Line, $^a$ Absorption Line, $^c$ Complex Line Profile, $^P$ P Cygni Profile, $^p$ Minor P Cygni Profile, $^i$ Minor Inverse P Cygni Profile}
\twocolumn

In some cases, we measure lines with complex features (e.g., P Cygni profile) due to a lack of pure emission/absorption lines. Lines used with complex features are marked in Table~\ref{tab:tab1}. These lines are used to measure the systemic velocity of targets because 1) Features are minor and the radial velocities are consistent with another pure emission/absorption line (e.g. R110 and MWC112). 2) For each target, radial velocity over several epochs remain constant (e.g., Sk-69$^\circ$271, R85, R71, S Dor/emission, and R143). However, in three cases (S Dor/absorption, R110 and Sk-69$^\circ$142a), the absorption velocities are highly variable either due to wind variability or binary motion. Both of these options are discussed in detail below.

\begin{figure}
\includegraphics[width=\linewidth]{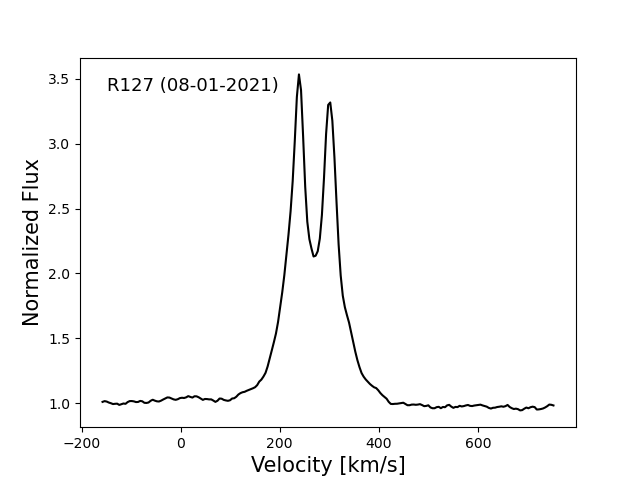}
\caption{Spectrum of R127 showing the nebular [N~{\sc ii}] $\lambda$6583 \AA \ line, which is emitted by the expanding shell nebula, and not the star's photosphere or wind. The two emission peaks trace blueshifted and redshifted sides of an expanding shell in a spectrum extracted at the position of the central star. We fit combined Gaussian profiles to these two peaks, the systemic radial velocity of R127 is taken to be mid-way between these two lines. }\label{fig:Forbidden}
\end{figure}

{\it Nebular lines:}  A few of our targets (R127, S119, S61, and R143) are surrounded by bright circumstellar shell nebulae at large distances from the star (i.e. more than 1 arcsec away, or  $R > 50,000$ AU).  Table~\ref{tab:nebula} presents the nebular radial velocities for these LBVs, measured using the [N~{\sc ii}] $\lambda$6583 \AA \ line. The center of expansion for shell nebulae is expected to give a good measure of the systemic velocity, as long as the star is at the kinematic center of the shell.  This appears to be the case for R127, which has an ellipsoidal nebula centered on the star, and for S61 which has an apparently spherical nebula centered on the star \citep{Stahl87}.  The other two LBVs, however, may not be at the kinematic center of their nebulae. HST and radio images show a very asymmetric nebula around R143, with the brightest emission concentrated on one side of the star \citep{A12}. Another exception might be if the star is moving rapidly compared to the surrounding interstellar medium (ISM), in which case the nebula might more closely resemble a bow shock, and its emission might trace gas that has been decelerated on one side compared to its initial ejection speed. This may be the case for S119 \citep{D01}, which turns out to be the fastest moving LBV in our sample. For LBV R127, most of the lines are doubled peaked due to a nebula. Therefore, we fit multiple Gaussians to the [N~{\sc ii}] $\lambda 6583$ \AA \ line to determine radial velocity. Fig.~\ref{fig:Forbidden} shows the double-peaked [N~{\sc ii}] emission lines in R127, which trace the front (blueshifted) and back (redshifted) sides of an expanding shell.  The systemic radial velocity of R127 is taken as halfway between the two lines. For LBV S61, the nebular velocities are consistent with emission velocities.

Radial-velocity estimates for each target are obtained by calculating an average of all individual measurements using the same method (absorption, emission, or nebular). Each individual radial velocity estimate is corrected for the motion of the Earth at the time of observation using the IRAF task rvcorrect, in order to put the velocities in a heliocentric reference frame. The measurement errors are also combined in quadrature to estimate the error of multiple epochs. In general, we find that absorption velocities are usually blueshifted compared to emission due to wind absorption, as noted above. For a few sources, we find not only that the absorption velocities differed from emission velocities, but that the absorption velocities also changed with time.  We attribute this mainly to Doppler shifts from orbital motion in binaries rather than wind variability.  We discuss this in detail in section~\ref{sec:abslines}, where we also investigate the photometric variability during the time frame when our spectra were obtained. 

\begin{table}
\caption{Radial Velocities using nebular [N~{\sc ii}] $\lambda$6583 \AA \ line. Some of the uncertainties are less than 0.05 but we round them up to 0.1.}
\begin{center}
\begin{tabular}{ l c c }
\hline
Name&Obs.\ Date & Radial Velocity [km s$^{-1}$] \\
\hline
\hline
R127 & 10-25-2017& 266.4$\pm$0.1\\
 & 03-12-2018& 265.9$\pm$0.1\\
 &  09-17-2018& 266.4$\pm$0.1\\
 & 04-24-2021& 268.1$\pm$0.1\\
 & 04-25-2021& 268.7$\pm$0.1\\
  & 08-01-2021& 269.6$\pm$0.1\\
\hline
R143 & 04-23-2021& 271.0$\pm$0.1\\
 & 04-25-2021& 268.3$\pm$0.1\\
  & 08-01-2021& 268.8$\pm$0.1\\
  
  \hline
  (S119) & 03-12-2018& 156.7$\pm$0.1\\
 & 04-24-2021& 156.3$\pm$0.2\\
 & 04-25-2021& 158.9$\pm$0.1\\
  & 08-01-2021& 161.3$\pm$0.1\\
  
    \hline
 (S61) & 04-23-2021& 286.2$\pm$0.1\\
 & 04-25-2021& 285.9$\pm$0.1\\
  & 08-01-2021& 286.0$\pm$0.1\\
 \hline
\label{tab:nebula}
\end{tabular}
\end{center}
\end{table} 

\begin{figure}
    \includegraphics[width=0.45\textwidth]{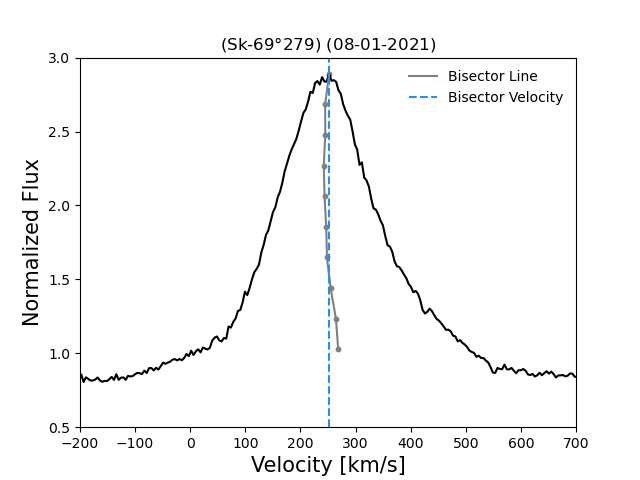}\par 
    \includegraphics[width=0.45\textwidth]{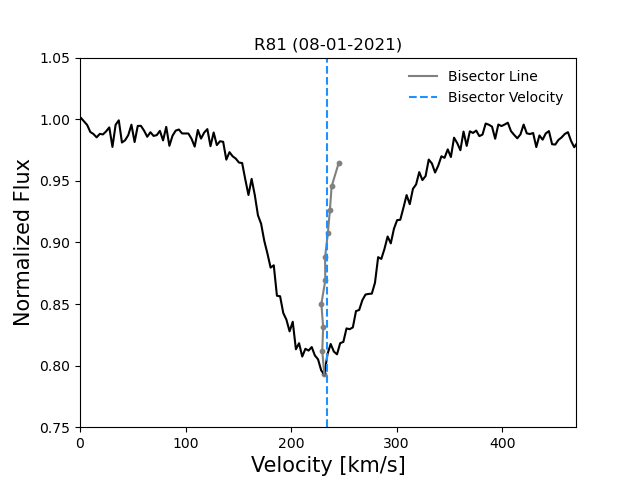}\par 
\caption{The top panel shows an example of an emission line (H$\alpha$ $\lambda$6563 \AA) used to estimate the bisector velocity. The bisector of the line profile is shown in grey points. The bottom panel shows an example of an absorption line (He~{\sc i} $\lambda$4388 \AA) and its corresponding bisector values. The bisector values are midway between the wings of the line profile and the bisector
velocity (dashed blue line) is the average of bisector points. Heliocentric bisector velocities are listed in Table~\ref{tab:bisector}.}\label{fig:bisector}
\end{figure}

To verify the radial velocity estimates from Gaussian fittings, we also estimate the radial velocities of a couple of targets using the bisector method.  We estimate the line bisector using the spectral analysis tool provided by the Esac Science Data Center (VOSpec)\footnote{\url{http://www.cosmos.esa.int/web/esdc/vospec}}. Fig.~\ref{fig:bisector} show bisector values for  H$\alpha$ and He~{\sc i} lines. The grey points show the bisector of the line profile, which is made up of the midpoints of the horizontal lines connecting the wings of the line profile. The bisector velocity of a spectral line is the average of the midpoints. Table~\ref{tab:bisector} shows the radial velocity of a couple of targets using the bisector method, which are in good agreement with the radial velocity estimated using a Gaussian fitting (listed in Table~\ref{tab:tab1}). The uncertainties listed in the table correspond to the standard error of the mean.

In addition to the bisector method, we estimate the flux weighted centroid of a couple of lines as well. The flux weighted velocities are consistent with the Gaussian fitting method used in this work ($\sim$ within 1-2 km s$^{-1}$). The flux weighted method might be more reliable than the bisector method because the centroid is weighted by the stronger part of the line, but the bisector line might get skewed by the faint line wing. However, we emphasize that small differences in radial velocity estimates using different methods are insignificant because they do not affect the inferred velocity dispersion of LBVs.

\begin{table}
\caption{Examples of radial velocities estimated using the bisector method. Bisector velocities are in good agreement with velocities estimated by fitting a Gaussian profile to the spectral line (shown in Table~\ref{tab:tab1}).}
\begin{center}
\begin{tabular}{ l c c }
\hline
Name&Obs.\ Date & Radial Velocity [km s$^{-1}$] \\
\hline
\hline
R81 & 08-01-2021& 256.5$\pm$4.1$^{e}$\\
 & & 233.9$\pm$3.8$^{a}$\\
\hline
Sk-69$^\circ$142a & 04-25-2021& 261.6$\pm$1.9$^{e}$\\
 & & 195.9$\pm$1.6$^{a}$\\
  
  \hline
(Sk-69$^\circ$279)  &  08-01-2021& 251.2$\pm$2.8$^{e}$\\
 & & 223.3$\pm$3.3$^{a}$\\
 \hline
\label{tab:bisector}
\end{tabular}
\end{center}
\footnotesize{$^e$ Emission Line, $^a$ Absorption Line}
\end{table}

\begin{figure}
\includegraphics[scale=0.5,angle=0]{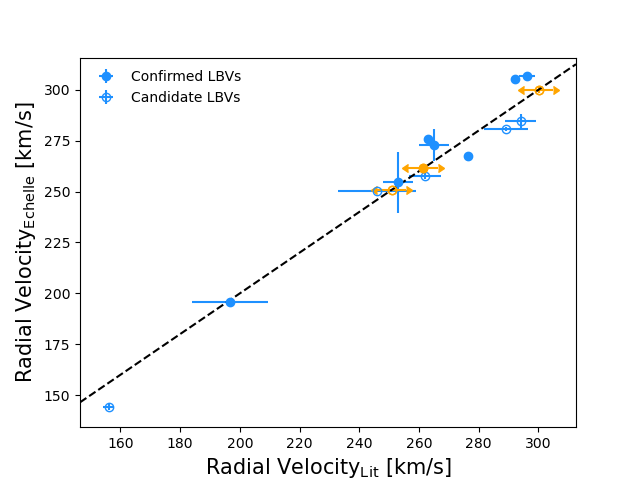}
\caption{Heliocentric emission radial velocities measured using MIKE echelle spectra compared to emission radial velocities reported previously in the literature (see Table~\ref{tab:tab2}). Filled circles represent the LBVs and open circles represent the LBV candidates. Orange circles represent Sk-69$^\circ$142a, Sk-69$^\circ$279, and Sk-69$^\circ$271, which have no published literature radial velocities using emission lines. Some literature values are slightly different than this work because the authors used lines with P Cyg features, or authors noted that the quality of their radial velocity estimates are low (see comments in Table~\ref{tab:tab2} for more details). In most cases, echelle error bars are smaller than points. While in several cases, echelle spectra provide smaller uncertainty than previous estimates in the literature, a couple of error bars are larger because they are a result of several individual errors combined in quadrature to estimate the error of multiple epochs.}
\label{fig:EchellevsLitEmission}
\end{figure}

\begin{figure}
\includegraphics[scale=0.5,angle=0]{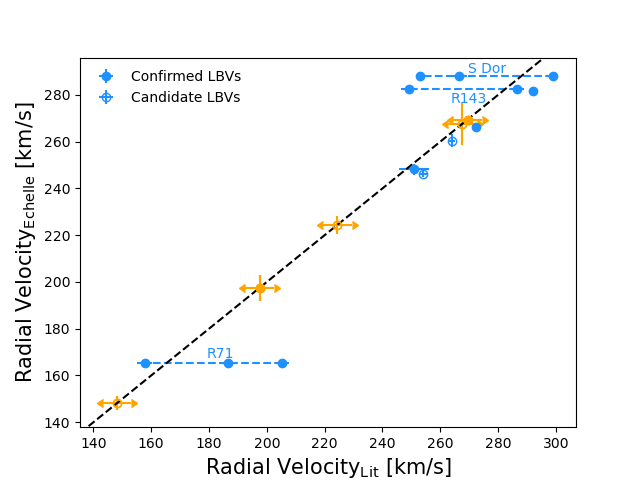}
\caption{Same as Fig.~\ref{fig:EchellevsLitEmission}, but using absorption lines. The orange circles represent R110, Sk-69$^\circ$142a, S61, S119, and Sk-69$^\circ$279, which do not have published radial velocities using absorption lines.}\label{fig:EchellevsLitAbsorption}
\end{figure}

We also consider how our new radial velocity measurements compare with previously published velocities for the same LBV targets.  Table~\ref{tab:tab2} presents the literature radial velocities for each LBV along with the corresponding references. For LBVs with observations at multiple epochs, we report the mean of the available velocities, and the errors are combined in quadrature. Figs.~\ref{fig:EchellevsLitEmission} and \ref{fig:EchellevsLitAbsorption} compare LBV radial velocities that we measure using MIKE echelle spectra to these literature values. Overall, the literature radial velocities are
consistent with the radial velocity estimates in this work, although in several cases, the new echelle spectra provide smaller uncertainty. 

\begin{table*}
\caption{Previous literature radial velocities for LMC LBVs and candidate LBVs (in parentheses).}
\begin{adjustbox}{max width=\textwidth,center}
\renewcommand{\arraystretch}{2.}
\begin{tabular}{llcll} 
\hline
Name          &Line ID  &Radial Velocity$_{\rm Lit}$ [km s$^{-1}$]  &Reference & Comment \\
\hline\hline
R127& Multiple lines&  276.5$\pm0.4^e$ & \cite{M09}  &  P Cyg lines used \\
\hline
  S Dor& Multiple lines& 296.2$\pm2.6^e$ &  \cite{M09} & Avg.\ of 3 epochs  \\
 & Multiple lines& 272.8$\pm2.8^a$ &  \cite{M09}  & Avg.\ of 3 epochs,  P Cyg lines used  \\

\hline
R81 & S~{\sc ii}, [N~{\sc ii}], [Ti~{\sc ii}] & 253.0$\pm5.0^e$& \cite{T02} &   \\
 & Ca~{\sc ii} & 251.0$\pm5.3^a$& \cite{A72} &  \\
\hline
R110 & [Fe] and [N~{\sc ii}] & 265.0$\pm5.0^e$&  \cite{C18} &  P Cyg lines used\\
\hline
R71 & Multiple lines &196.6$\pm12.7^e$&  \cite{M09} & Avg.\ of 3 epochs \\
 & Multiple lines &183.3$\pm4.0^a$&  \cite{M09} &  Avg.\ of 3 epochs\\
\hline
MWC112 & Compared with ref. spectra &  272.3$\pm$1.2&  {\it Gaia} DR2 & \\
\hline
R85 & Multiple lines &  292.0& \cite{F60} & Low quality estimate \\
\hline
R143& Multiple lines & 263.0&  \cite{F60} & Low quality estimate \\
& Multiple lines & 267.9$\pm3.5^a$& \cite{M09} & Avg.\ of 2 epochs \\
\hline
Sk-69$^\circ$142a & N/A& N/A &  N/A & \\
\hline
(R84) & He~{\sc i}, He~{\sc ii}, N~{\sc iii}, H$\alpha$ & 246.0$\pm13.0^e$ & \cite{N96} &  \\
 & Multiple lines & 254.2$\pm1.2^a$ & \cite{M09} &  \\
\hline
(R126) & Balmer lines, Fe~{\sc ii}, [Fe~{\sc ii}], Ca~{\sc ii k} & 262.0$\pm5.3^e$& \cite{A72} & \\
\hline
(S119) & [N~{\sc ii}] & 156.0$\pm2.0^e$& \cite{W03} &   \\
\hline
(Sk-69$^\circ$271) & He~{\sc i} and Si~{\sc iii} & 264.0$\pm$1.2$^a$&  \cite{E15} & \\
\hline
(R99)& Multiple lines &289.0$\pm7.3^e$& \cite{M09} & Avg. of 3 epochs \\
\hline
(S61)& H$\beta$ to H~{\sc $\epsilon$}, He~{\sc i}, He~{\sc ii}, N~{\sc iii}, Si~{\sc iii} & 294.0$\pm5.3^e$& \cite{A72} & \\
\hline
(Sk-69$^\circ$279) & N/A & N/A &   N/A & \\
\hline
\label{tab:tab2}
\end{tabular}
\end{adjustbox}
\footnotesize{$^e$ Emission Line, $^a$ Absorption Line}
\end{table*}

Some radial velocity estimates for emission lines in the literature are slightly different than radial velocity estimates in this work.  This can be due to multiple reasons: 1) Some authors use lines with P Cyg features (e.g., R127 in \citet{M09}), 2) Some radial velocity estimates are marked as low quality by the authors (e.g., R85 in \citet{F60}), or  3) Intrinsic line variation due to wind variability or binarity may also cause variation in the radial velocity estimates. 

\subsection{Absorption Line Variability}\label{sec:abslines}
In this work as well as previous estimates in the literature, the absorption velocities for some LBVs are highly variable with time. In principle, this may be caused by either orbital motion in binaries, or by changes in an LBV's wind velocity with time.  In the hot quiescent phases of LBVs, their photospheric radii are expected to be relatively small and their wind velocity is relatively fast. Then, during their outburst, the mass-loss rate may increase, the photospheric radius increases, and the wind velocity may decrease \citep{s20}. If an LBV's wind is dense enough, this could potentially lead to the formation of the pseudo-photosphere. If absorption lines are formed in the pseudo-photosphere, the change in the wind velocity can potentially cause apparent variation in the measured radial velocities in lines affected by the wind absorption.  
 
Table~\ref{tab:tab1} shows that the LBVs R110, R81, S Dor, Sk-69$^\circ$142a and the LBV candidate Sk-69$^\circ$279 have highly variable absorption velocities. To test whether the radial velocity variability is related to the wind variability or binary orbital motion, we estimate the line width of H$\alpha$ emission. Table~\ref{tab:gfwhm} shows Gaussian full-width at half-maximum (GFWHM) values of H$\alpha$ for the five LBVs with variable absorption velocities. For all five LBVs, the wind speed inferred from H$\alpha$ does not change enough to explain the changes in the radial velocity seen in absorption lines.  Wind variability is therefore probably not the primary cause of absorption velocity changes with time. 

\begin{table}
\caption{Gaussian full-width at half-maximum (GFWHM) of H$\alpha$ emission lines measured using echelle spectra of LMC LBVs. Some uncertainties are less than 0.05 but we round them up to 0.1.}
\begin{center}
\begin{tabular}{ l c c }
\hline
Name&Obs.\ Date & GFWHM [km s$^{-1}$] \\
\hline
\hline
R110 & 10-25-2017& 73.3$\pm$0.1\\
 & 04-25-2021& 76.1$\pm$0.1\\
 & 08-01-2021& 76.8$\pm$0.1\\
 & 10-02-2021& 73.4$\pm$0.1\\
 & 10-12-2021& 72.0$\pm$0.1\\

\hline
R81 &03-11-2018& 132.4$\pm$0.1\\
 & 04-24-2021&  121.8$\pm$0.1\\
 & 08-01-2021& 129.2$\pm$0.1\\
  & 10-02-2021& 124.5$\pm$0.1\\
 & 10-12-2021& 129.5$\pm$0.1\\

\hline
S Dor &03-11-2018&  49.8$\pm$0.1\\
  & 04-24-2021&   44.7$\pm$0.1\\
  & 08-01-2021& 48.6$\pm$0.1  \\
   & 10-02-2021& 39.7$\pm$0.1\\
 & 10-12-2021& 38.9$\pm$0.1\\

 \hline
Sk-69$^\circ$142a &03-12-2018& 58.1$\pm$0.1\\
 & 04-25-2021& 77.2$\pm$0.1\\
& 08-01-2021& 61.3$\pm$0.1 \\
 & 10-02-2021& 56.6$\pm$0.1\\
 & 10-12-2021& 84.5$\pm$0.1\\
\hline
(Sk-69$^\circ$279)&  10-26-2017 & 231.6$\pm$0.5\\
& 04-24-2021&  219.9$\pm$0.3\\
&  08-01-2021&  222.5$\pm$0.2\\
 & 10-02-2021& 208.8$\pm$0.2\\
 & 10-12-2021& 204.4$\pm$0.4\\
\hline
\label{tab:gfwhm}
\end{tabular}
\end{center}
\end{table}

Using photometric data from the All Sky Automated Survey for SuperNovae \citep[ASAS-SN;][]{S14, K17} and the American Association of Variable Star Observers (AAVSO) International Database\footnote{\url{https://www.aavso.org/data-download}}, we also test whether the radial velocities of LBVs may be correlated with apparent magnitude changes with time.  This should be the case if their velocities are influenced by eruptive wind variability. If the apparent magnitude does not show significant variation ($>$ 0.5 mag) with time, this suggests that these LBVs are not transitioning from quiescence to outburst state or vice versa. Below, we explore the light curves for the five LBVs that exhibit significant absorption radial velocity changes, in order to further confirm that the variation in the absorption velocities is due to orbital motion in binaries rather than wind variability.

\begin{figure}
\includegraphics[scale=0.5,angle=0]{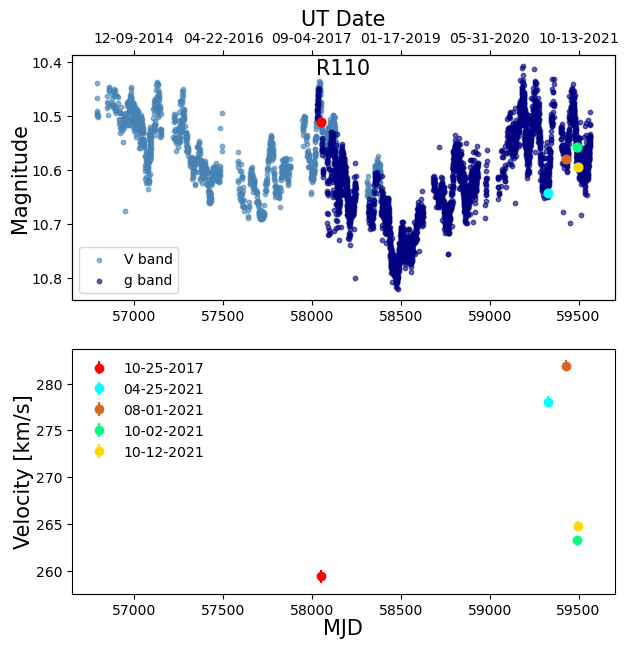}
\caption{The top panel shows the ASAS-SN light curve of R110. Photometry coincident in time with our MIKE spectra are labeled as five large filled circles. The magnitude error bars in the light curves shown above are smaller than the points. Microvariations and S Dor activity are evident in the light curve. The bottom panel shows the absorption radial velocity of R110 over time.}\label{fig:R110LC}
\end{figure}

{\it R110}: The top panel in Fig.~\ref{fig:R110LC} shows the ASAS-SN light curve of R110 over the past 7-8 yr. The light curve shows microvariations on top of the S Dor activity. We fold the light curve using different periods and find that there is no clear period matching the oscillations, indicating that the R110's microvariation is not regular. \citet{V98} studied the photometry related to the microvariations of R110 and also found that these oscillations have various time scales between 50d to 100d. While the photometric variations are not strictly periodic and may not be caused by a binary, Fig.~\ref{fig:R110LC} also shows that the $V$- and $g$-magnitude of R110 exhibits no major eruptions (observed variations are only about 0.2 mag) that might substantially alter the nature of the wind. This agrees with the relatively constant GFWHM of H$\alpha$ noted above.  This implies that the variation in the absorption velocities (shown in bottom panel) is not a result of wind variability and is likely related to binary motion instead.

\begin{figure}
\includegraphics[scale=0.5,angle=0]{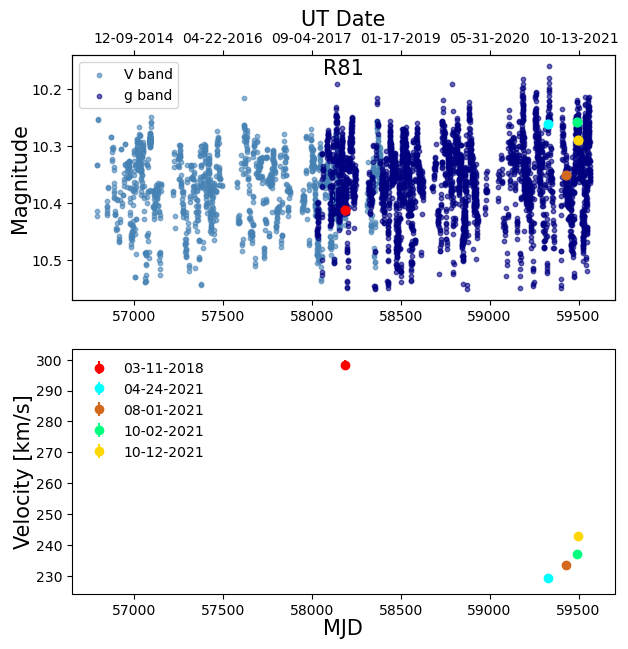}
\caption{The top panel shows the ASAS-SN  light curve of R81. Photometry coincident in time with our MIKE observations are labeled as large filled circles. The bottom panel shows the absorption radial velocity of R81 over time.}\label{fig:R81LC}
\end{figure}

\begin{figure}
\includegraphics[width=\linewidth,scale=0.46,angle=0]{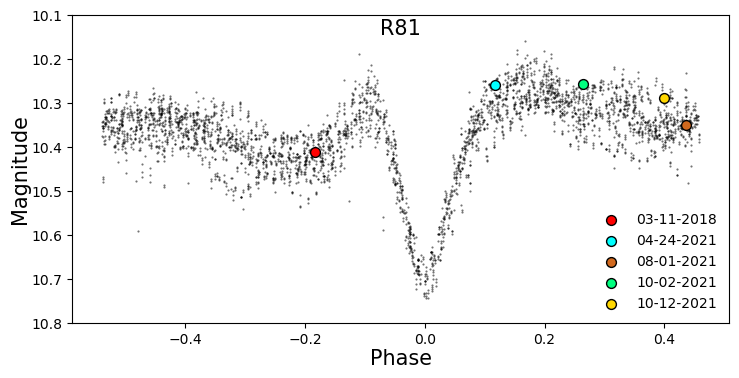}
\caption{Phase diagram showing the folded light curve of R81 with the period of 74.566 days recommended by \citet{S87}. Photometry coincident in time with our MIKE observations are labeled as colored circles with black outlines.}\label{fig:R81LC-folded}
\end{figure}

{\it R81}: The light curve of R81 is shown in the top panel of Fig.~\ref{fig:R81LC}. We find that the $V$- and $g$-magnitudes of R81 stay roughly the same over various epochs, indicating no eruptive outburst variability. However, R81 is a known eclipsing binary \citep{S87} with an orbital period of 74.566 days \citep{T02}.  Fig.~\ref{fig:R81LC-folded} shows all ASAS-SN data folded with that orbital period. In addition to the primary minimum, there is another minimum around phase $-$0.25. If this minimum is due to a secondary eclipse, the system is eccentric. While the line width of H$\alpha$ emission does not change over time and the velocities of Si~{\sc ii} emission do not change significantly, the bottom panel of Fig.~\ref{fig:R81LC} shows that He~{\sc i} absorption velocities are significantly redshifted at the one 2018 epoch and blueshifted at four epochs in 2021.  We conclude that the H$\alpha$ and Si~{\sc ii} emission does not change because the wind speed remains constant; the period is fairly short, and so H$\alpha$ forms at a large radius or it is the average over the whole orbit. He~{\sc i} absorption velocities, on the other hand, might be dominated by one of the stars' photospheres and thus show larger changes in velocity due to orbital motion.  We note that the one redshifted He~{\sc i} velocity in 2018 occurred before the primary eclipse (phase $\sim$-0.2 in Fig.~\ref{fig:R81LC-folded}, and the four blueshifted He~{\sc i} velocities were after the primary eclipse.  This means that the He~{\sc i} absorption is associated with the hotter companion star that is behind during the primary eclipse. \cite{T02} derived stellar parameters of T$_{eff}$=19500 K, R=96 R$_{\sun}$ and T$_{eff}$=8000 K, R=58 R$_{\sun}$ for the primary and secondary star respectively. This suggests that He~{\sc i} absorption is probably arising from the LBV, which is behind during the primary eclipse.  

\begin{figure}
\includegraphics[scale=0.5,angle=0]{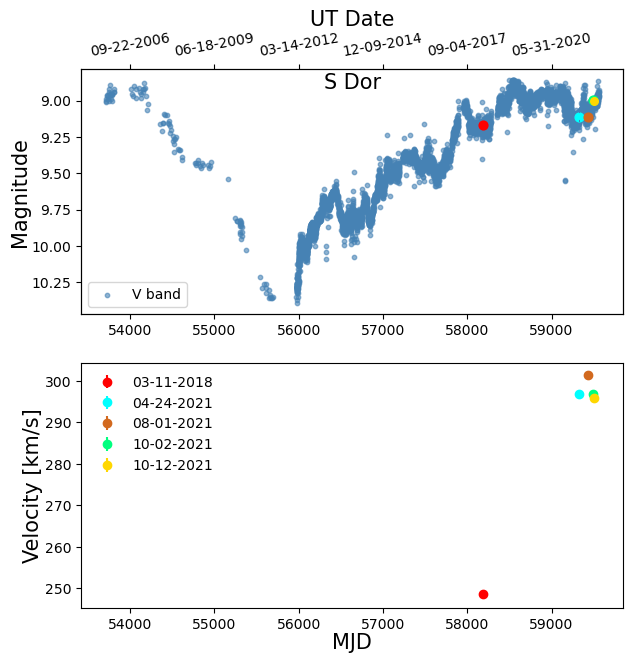}
\caption{The top panel shows the AAVSO light curve of S Dor. Photometry coincident in time with our MIKE observations are labeled as large filled circles. The bottom panel shows the absorption radial velocity of S Dor over time.
}\label{fig:SdorLC}
\end{figure}

{\it S Dor}: The top panel in Fig.~\ref{fig:SdorLC} shows that the $V$ magnitude of S Dor does not change much during the time of the MIKE observations. Although S Dor was experiencing an S Dor eruption during the time when our MIKE spectra were obtained, this was a long-lasting (several-year) eruptive phase, and so all our spectra were obtained when S Dor was in a similar maximum state.  Thus, while the wind of S Dor may be significantly altered from its quiescent state during our observations, we would not expect much difference in wind properties from one of our epochs of MIKE spectra to the next, and indeed, we find little variation in S Dor's H$\alpha$ wind emission as noted above.  We, therefore, conclude that the variations in Si~{\sc ii} absorption velocities (see the bottom panel of Fig.~\ref{fig:SdorLC}) that we detect are caused by binary orbital motion.

\begin{figure}
\includegraphics[scale=0.5,angle=0]{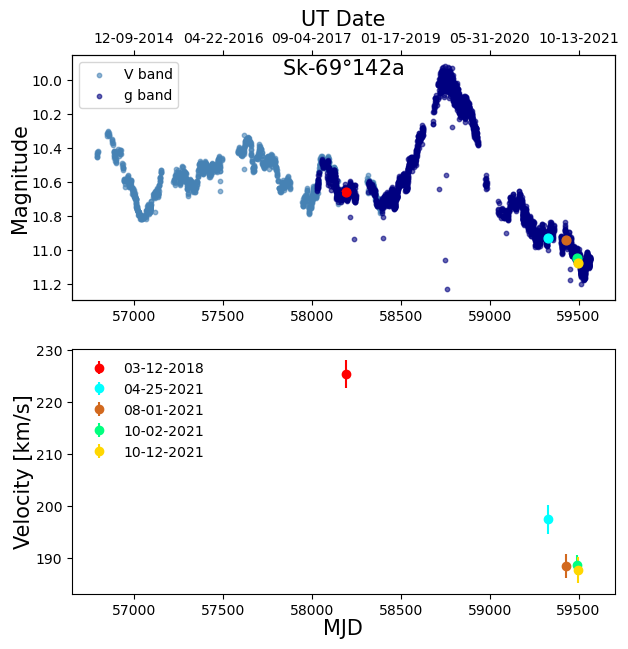}
\caption{The top panel shows the ASAS-SN  light curve of Sk-69$^\circ$142. Photometry coincident in time with our MIKE observations are labeled as large filled circles. Sk-69$^\circ$142a experienced its second known outburst in late 2019, confirming that it is an LBV and no longer just a candidate. The bottom panel shows the absorption radial velocity of Sk-69$^\circ$142a over time.}\label{fig:SK-69142aLC}
\end{figure}

{\it Sk-69$^\circ$142a}:  This star was previously considered to be an LBV candidate.  However, it recently had an outburst in 2013 (see below), and examining the ASAS-SN light curve shown in Fig.~\ref{fig:SK-69142aLC} (top panel), we find that it experienced another major outburst in 2019/2020.  This is the second documented outburst of Sk-69$^\circ$142a, confirming its shift in status from a candidate to a confirmed LBV.  \citet{SO86} classified  Sk-69$^\circ$142a as an S Dor variable and \citet{W17} documented its first outburst in 2013. On 2019 Sep 21, Sk-69$^\circ$142a reached a $g$-magnitude of 9.9, which is roughly as bright as its first outburst ($V$=9.8 mag). Even though Sk-69$^\circ$142a experienced a recent outburst, Fig.~\ref{fig:SK-69142aLC} shows that our MIKE observations were all obtained at relative quiescence (before and after the outburst, but not during it) when the apparent magnitude was approximately the same\footnote{In other words, we, unfortunately, missed getting spectra in outburst.}.  Thus, eruptive wind variability is not likely to explain radial velocity changes in our spectra (bottom panel), and again, these are likely attributable to binary orbital motion. So far, there has been no published systematic search for companion of Sk-69$^\circ$142a. 

\begin{figure}
\includegraphics[scale=0.5,angle=0]{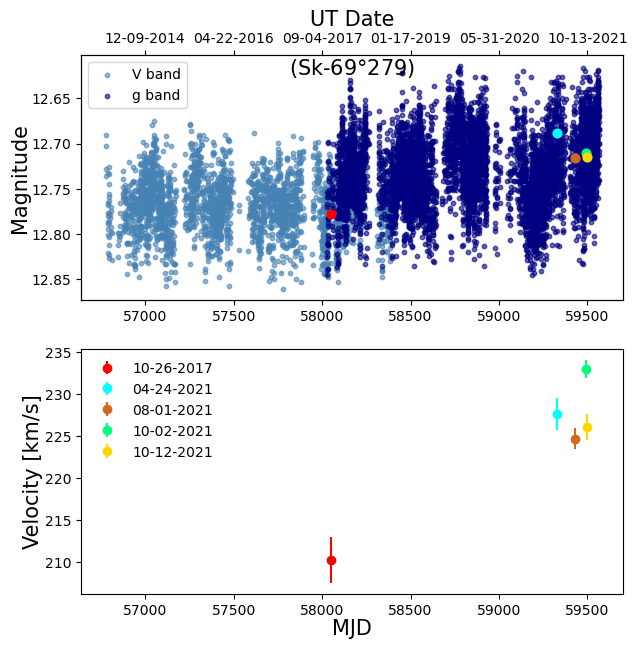}
\caption{The top panel shows the ASAS-SN  light curve of Sk-69$^\circ$279. Photometry coincident in time with our MIKE observations are labeled as large filled circles. The observed variation in the light curve is consistent with noise (and perhaps a slight shift due to the change in the filter) and does not indicate any significant eruptive variability or eclipses. The bottom panel shows the absorption radial velocity of Sk-69$^\circ$279 over time.}\label{fig:SK-69279LC}
\end{figure}

{\it Sk-69$^\circ$279}: This target is an LBV candidate without any outburst observed yet. The ASAS-SN  light curve is shown in the top panel of Fig.~\ref{fig:SK-69279LC}, and it reveals no major magnitude change (amplitude less than about 0.1 mag) with time that would be evidence for significant eruptive wind variability.  Therefore, any radial velocity changes (see the bottom panel of Fig.~\ref{fig:SK-69279LC}) in our spectra are probably not caused by wind variability, and we suggest that the absorption-line velocity changes are instead attributable to binary orbital motion. This is the first indication that Sk-69$^\circ$279 is in a binary system.

Above, we showed that all five LBVs R110, R81, S Dor, Sk-69$^\circ$142a and the LBV candidate Sk-69$^\circ$279 with highly variable absorption velocities have roughly constant apparent magnitudes over time (or at least, that they do not show substantial photometric changes indicative of major shifts in eruptive status between spectral epochs), and their wind speeds indicated by the H$\alpha$ width do not change significantly. This implies that the absorption velocity changes are most likely due to the binary motion. This is perhaps not surprising, as \citet{mahy2022} found a high current binary fraction among LBVs in the Milky Way.  S Dor and R110 have known companions detected by the Fine Guidance Sensor on the Hubble Space Telescope. \cite{A14} found that R110 and S Dor's companion are $>$0.09\arcsec and 1.7\arcsec away respectively. In both cases, companion stars are widely separated ($\sim>$4,500 and 85,000 AU) and they can't cause the observed radial velocity change. A closer look is needed to find the companions that are causing the radial velocity change. R81 is also known as a binary because it has been classified as an eclipsing binary by \citet{S87}.  For  Sk-69$^\circ$142a and Sk-69$^\circ$279 this is the first indication to our knowledge that these stars are in binary systems.

While binary motion may affect the radial velocity estimates using absorption lines, it might not strongly affect the emission velocities. Emission lines are formed at different radii in the wind; for example, H$\alpha$ can be emitted from a large volume beyond the orbit of the binary system, and it may trace the average velocity of wind material emitted over several orbits. Therefore, one can observe more highly variable radial velocities over time using absorption lines formed in the stellar atmospheres. This suggests that the emission velocities are better estimates for the systemic velocity of R110, R81, S Dor, Sk-69$^\circ$142a, and Sk-69$^\circ$279.  We expected this anyway for most of the sample since absorption lines tend to be slightly blueshifted compared to emission lines because some absorption occurs in the outflowing wind. In the next section, using the radial velocity of LMC LBVs and Bayesian inference, we infer the velocity dispersion of LBVs to constrain whether LMC LBVs received significant kicks from their companion's SN explosion.
 
\section{VELOCITY DISPERSIONS}\label{sec:dispersions}
In this section, we describe a method to infer the velocity dispersion of LBVs using systemic radial velocities and Bayesian inference. To infer the velocity dispersion of LBVs, we must first correct the observed radial velocities for the projected velocity due to the rotation of the galaxy at each location. Then, we compare the velocities of LBVs with other evolved massive stars like RSGs to determine if their kinematics are unusual. In the following sections, we first describe a rough method to estimate the velocity dispersion of LBVs, and then we infer a more accurate velocity dispersion of LBVs using Bayesian inference.

\begin{figure*}
\centering
\includegraphics[width=0.7\textwidth]{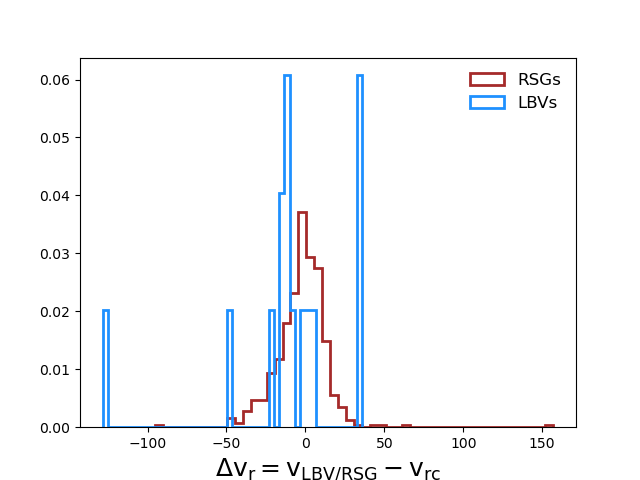}
\caption{Histogram of $\Delta \rm{v_r}$, the difference between the observed radial velocities and the projected line-of-sight velocities derived from the rotation curve model by \citet{C22}. The blue histogram shows the distribution for LBVs measured from their emission line velocities, and the red histogram shows the distribution for RSGs. The RSGs histogram shows an example of a population of evolved massive stars that includes both single and binary stars.}
\label{fig:LBVvsRSG}
\end{figure*}

\subsection{A Rough Statistical Inference}
\label{sec:statistical inference}
In order to measure the velocity dispersion of LBVs, we must first compare the observed radial velocities of LBVs to their local environments, in order to correct for projected velocities due to the LMC's rotation. In principle, the 21-cm line of neutral atomic hydrogen can be used as a probe of gas along the line of sight to trace massive stars' environments. 
A star cluster formed out of a cloud should inherit the centroid velocity of the cloud, with some random velocity dispersion from their birth cluster.  Since massive stars have short lifetimes, we might expect there to be a good general agreement between the local gas and massive stars. Therefore, initially, we aimed to compare each LBV's velocity to the local H~{\sc i} gas. However, recent evidence of multiple peak velocity profiles is obtained using high-resolution H~{\sc i} map \citep{K98,S03}. The complex multi-component velocities of H~{\sc i} in the south-east region of LMC might be due to gas streams \citep[including recent interaction with the SMC;][]{B12,B16}. The H~{\sc i} profiles with multiple peaks can cause confusion along each line of sight, with differences in individual H~{\sc i} component velocities that are comparable to or larger than the dispersion of stellar velocities. 

Instead of using H~{\sc i} data to sample LMC rotation, we use the projected line-of-sight velocities derived by fitting a model of internal rotation of the LMC to the observed kinematics of RSGs \citep{C22}. Briefly, we fit a disk rotation model to $\sim$1,000 RSGs with proper motions from Gaia EDR3 \citep{GaiaEDR3} and line-of-sight velocities from the Hydra-CTIO observations \citep[Olsen et al. in prep.]{O11}. Our modeling procedure jointly fits twelve parameters: the location of the kinematic center on the sky, the bulk transverse motion along the RA and Dec axes, the line-of-sight velocity of the kinematic center, the position angle of the line of nodes, the inclination of the disk, two parameters describing the shape and amplitude of the internal rotation curve, and the velocity dispersion in three orthogonal directions. We assume the distance to the LMC center to be 50.1~kpc and assume there is no precession or nutation in the LMC disk. 

For each LBV and LBV candidate in our sample, we calculate the difference in radial velocity (which we denote hereafter as $\Delta \rm{v_r}$) between the LBV's observed radial velocity from spectra (v$_{\rm{LBV}}$) and the projected rotation curve velocity at that same position (v$_{\rm rc}$). Table~\ref{tab:FinalVelocity} lists the v$_{\rm{LBV}}$, v$_{\rm rc}$, and $\Delta \rm{v_r}$ for each target.  The resulting values of  $\Delta \rm{v_r}$ provide a measure of how much LBVs differ from the expected average kinematics of their local environments. Appendix \ref{sec:appendix} shows the observed radial velocity compared to the local rotation curve velocity for each target. We do the same comparison for a sample of LMC RSGs, using radial velocities reported by \citet{N12}.  Fig.~\ref{fig:LBVvsRSG} shows histograms of the distribution of $\Delta \rm{v_r}$ values for LBVs (blue) compared to a sample of RSGs (red).  These are described in more detail in the following subsections.

\begin{table}
\caption{List of the LBVs and candidate LBVs (names in parentheses), the average  radial velocities (v$_{\rm{LBV}}$) derived from the emission lines (not absorption lines or strong P Cygni profiles), the local rotation curve velocities (v$_{\rm rc}$), and the difference  between the LBV's observed radial velocity and the projected rotation curve velocity at that same position ($\Delta \rm{v_r}$). The first round of uncertainties correspond to the Gaussian measurement errors (listed in Table~\ref{tab:tab1}) combined in quadrature to estimate the error of multiple epochs, while the second round of uncertainties correspond to standard error of the mean velocity.}
\begin{center}
\begin{adjustbox}{max width=\textwidth,center}
\begin{tabular}{lccc}
\hline
Name & v$_{\rm{LBV}}$ & v$_{\rm rc}$ & $\lvert\Delta \rm{v_r} \rvert$ \\
     & [km s$^{-1}$]  & [km s$^{-1}$]  & [km s$^{-1}$] \\
\hline
\hline 
R127 & 267.5$\pm0.2\pm$0.6&266.1&1.4\\
S Dor & 306.8$\pm0.7\pm$0.6&270.7&36.1\\
R81 & 254.6$\pm15.0\pm$0.6&277.5&22.9\\
R110 & 273.0$\pm7.9\pm$1.6&273.2&0.2\\
R71 & 195.8$\pm0.3\pm$0.8&243.2&47.4\\
R85 & 305.1$\pm0.1\pm$0.6&270.5&34.6\\
R143 & 275.7$\pm1.5\pm$0.5&271.1&4.6\\
Sk-69$^\circ$142a & 261.5$\pm3.0\pm$0.8&274.3&12.8\\
(R84) & 250.5$\pm0.4\pm$1.4&266.6&16.1\\
(R126) & 257.7$\pm1.4\pm$0.7&267.8&10.1\\
(S119)& 144.0$\pm1.5\pm$0.4&272.5&128.5\\
(Sk-69$^\circ$271)&300.0$\pm1.4\pm$0.5&264.0&36.0\\
(R99)& 280.7$\pm1.1\pm$0.7&288.4&7.7\\
(S61)& 284.5$\pm3.6\pm$0.4&295.4&10.9\\
(Sk-69$^\circ$279)& 250.8$\pm0.2\pm$2.0&264.2&13.4\\
\hline
\label{tab:FinalVelocity}
\end{tabular}
\end{adjustbox}
\end{center}
\end{table}
\subsubsection{Velocity dispersion of RSGs}
To evaluate whether LMC LBVs on average are runaways or have been rejuvenated, one should compare their velocity dispersion with other evolved massive stars in the LMC that can serve as a control sample. One option is to compare LBV's velocity dispersion with O stars' velocity dispersion. However, RSGs are a better option because the total binary fraction of RSGs is lower than that of O stars. (O-type stars are still on the main sequence and their radial velocities are more strongly contaminated by short-period binaries, and while RSGs may have companions, these will be wide companions with low radial velocity shifts). Also, RSGs have more precise measurements of their radial velocities due to their narrow spectral lines. Moreover, like LBVs, RSGs are massive stars in a post-main sequence evolutionary stage. \cite{N12} identified 505 RSGs in the LMC with precise radial velocities. Therefore, in this work, we use RSGs as an example of evolved massive stars to compare with LBVs.

\begin{figure}
\includegraphics[scale=0.5,angle=0]{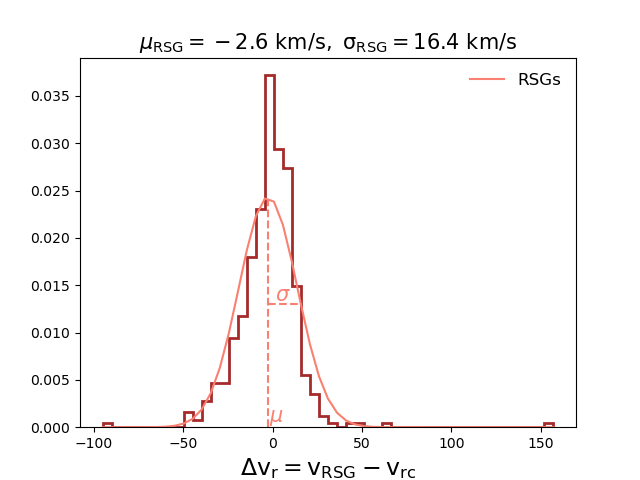}
\caption{Distribution of $\Delta \rm{v_r}$ values for RSGs, using observed radial velocities reported by \citet{N12}. The smooth red-orange curve shows a Gaussian fit to the RSG distribution, with centroid velocity ($\mu$) and velocity dispersion ($\sigma$) given at the top of the frame.}\label{fig:RSGRotationCurve}
\end{figure}

Fig.~\ref{fig:RSGRotationCurve} shows the histogram of the radial velocity of RSGs relative to their local rotation curve velocities, as well as a Gaussian fit to the distribution. The velocity dispersion of RSGs is found to be $\sim$ 16.4 km s$^{-1}$, which we take as representative of the velocity dispersion of evolved massive stars. The centroid of the velocity distribution of RSGs has a small blueshift of about $-$2.6 km s$^{-1}$. The dispersion of 16.4 km s$^{-1}$ implies a standard error of the mean of $\sim$0.7 km s$^{-1}$, which is $\sim$4 sigma significant.

\begin{figure}
\includegraphics[scale=0.5,angle=0]{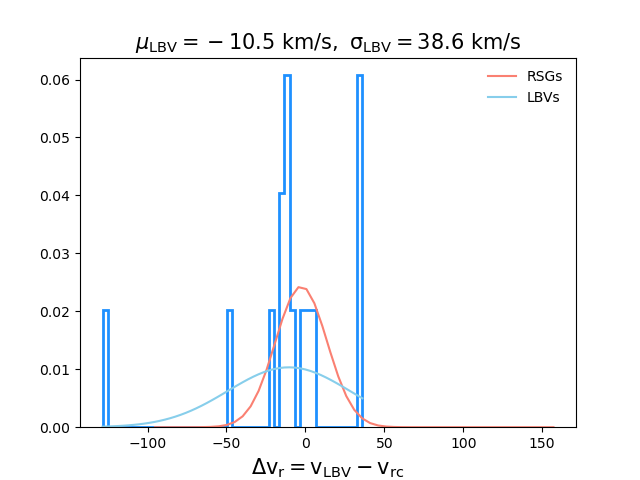}
\caption{Distribution of $\Delta \rm{v_r}$ values for LBVs calculated using radial velocities from emission lines. 
The smooth blue curve shows a Gaussian fit to the LBV distribution, with $\mu$ and $\sigma$ noted at the top as in Fig.~\ref{fig:RSGRotationCurve}. The red-orange curve reproduces the Gaussian fit to the RSG distribution from Fig.~\ref{fig:RSGRotationCurve}. LBVs clearly have a larger velocity dispersion than RSGs, and the centroid of their distribution also appears significantly blueshifted.} \label{fig:LBVRotationCurveEmission}
\end{figure}

\begin{figure}
\includegraphics[scale=0.5,angle=0]{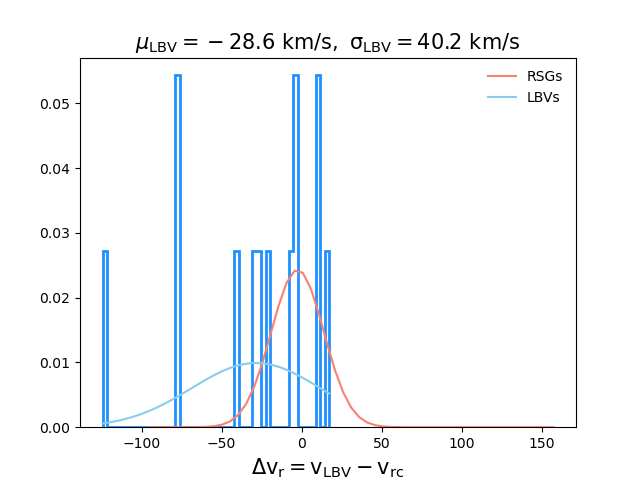}
\caption{Same as Fig.~\ref{fig:LBVRotationCurveEmission}, but showing $\Delta \rm{v_r}$ values for LBVs calculated using absorption lines.  The centroid of the LBV distribution is even more blueshifted than for emission lines.}\label{fig:LBVRotationCurveAbsorption}
\end{figure}

\subsubsection{Velocity dispersion of LBVs}
Figs.~\ref{fig:LBVRotationCurveEmission} and \ref{fig:LBVRotationCurveAbsorption} show the histograms of $\Delta \rm{v_r}$ for LBVs that are measured using emission and absorption lines, respectively.  Both methods indicate that the velocity dispersion of LBVs is $\sim$ 40 km s$^{-1}$. This is more than twice the measured velocity dispersion of RSGs. LBVs' higher velocity dispersion in comparison to RSGs may suggest that LBVs appear to be more isolated from other massive stars than expected because of unusual kinematics.

Fig.~\ref{fig:LBVvsRSG} shows the histogram of the relative velocity of LBVs using emission lines compared to RSG. Excluding one LBV (MWC112) that only has absorption lines in its spectrum, approximately 33\% of the LBVs in the LMC have velocities much larger than the velocities derived from the rotation curve model. This suggests that there is some additional intrinsic dispersion in the LBV distribution, as compared to RSGs. S119 has a velocity larger than 125 km s$^{-1}$ relative to the rotation curve model, which implies that S119 is a runaway LBV. This is supported by the detection of the bow shock in the Echelle observations by \citet{D01}. Four other LBVs ---  R71, S Dor, Sk-69$^\circ$271, and R85 --- have velocities that differ from the local rotation curve by more than 25 km s$^{-1}$, suggesting that these are also walkaways/slow runaways. If LBVs are born in a population that is slightly blueshifted (We discuss this in detail below), it would perhaps be more meaningful to compare LBV relative velocities to the centroid of their distribution. In this case, a few redshifted LBVs will have higher velocities, even though their radial velocities compared to the rotation curve are not high. This will lead to an even higher velocity dispersion of LBVs compared to RSGs, and a larger fraction of runaways. With these 5 out of 15 LBVs exhibiting high radial velocities, we measure a runaway fraction of around 33\%.  Note, however, that this is a lower limit because there could be additional runaway LBVs that show slow radial velocities if their motion is directed near the plan of the sky, resulting in small Doppler shifts despite relatively fast motion. With randomly oriented trajectories, we might expect an additional 2 or 3 runaway LBVs in this case, raising the runaway fraction to around 50\%. For comparison, the fraction of RSGs that have relatively high velocities above 25 km s$^{-1}$ is only 9\%.

Fig.~\ref{fig:LBVvsRSG} also shows that the centroid of the LBV distribution is skewed to blueshifted velocities with a larger offset than RSGs. In principle, this might arise due to the concentration of LBVs near 30 Dor, the influence of an LBV’s strong wind, or due to missing redshifted LBVs because of extinction. All of these options are discussed in detail in section~\ref{sec:centroid}. Given that there are uncertainties in estimating radial velocities, in the next subsection we infer the mean and the velocity dispersion of LBVs and RSGs through Bayesian inference.

\subsection{Bayesian Analysis}
\label{sec:Bayes}
To infer the velocity dispersion of LBVs, the posterior distribution for the model parameters, $\theta$ is the product of the likelihood $\mathcal{L}(data|\theta)$ and the prior probability $ P(\theta)$:
\begin{equation}
\label{eq:Bayes}
P(\theta|data) \propto \mathcal{L}(data|\theta)P(\theta)\, .
\end{equation}

\noindent The model parameters, $\theta$, are the mean velocity, $\mu$, and the velocity dispersion, $\sigma$. The observations, $data$, include radial velocities and the rotation curve velocities at the same location. To find the posterior distribution, equation~(\ref{eq:Bayes}), we choose uniform positive prior distributions for velocity dispersion, and the likelihood probability is: 
\begin{equation}
\mathcal{L}(data|\theta)= \frac{1}{\sqrt{2 \pi} \sigma_{\rm{v}}} \exp \left [\frac{-(\Delta \rm{v_r} -
  \mu)^2}{2{\sigma_{\rm{v}}}^2} \right ] \, ,
\end{equation}

\noindent where $\Delta \rm{v_r}$ is the observed radial velocity from spectra (v$_{\rm{LBV}}$ or v$_{\rm{RSG}}$) with respect to the modeled rotation curve velocity. $\sigma_{\rm {v}}$ is equal to $\sqrt{\sigma_{tot}^2+\sigma^2}$, where ${\sigma_{tot}}$ is the total uncertainty in estimating the radial velocity, and $\sigma$ is the velocity dispersion.

There are three sources of uncertainty in estimating the radial velocity of each LBV. One is the Gaussian measurement error, which is listed in Table~\ref{tab:tab1}. This is quite small, typically less than 1-2 km s$^{-1}$. Second, there is a source of error related to the spectroscopic wavelength calibration. The systematic uncertainty of wavelength calibration using the MIKE pipeline is smaller than the measurement errors and we ignore it in this analysis. As a test, we infer the velocity dispersion of LBVs by including a systematic uncertainty of around 1-2 km s$^{-1}$, and we find that the result is consistent with excluding this source of uncertainty. Third, there is a systematic uncertainty in estimating the average radial velocity of each target. For all the targets, the emission velocities remain approximately the same over time. Therefore, the variation between the individual radial velocity measurements is small ($\sigma<$ 5 km s$^{-1}$). However, the absorption velocities are sometimes  variable (up to 30 km s$^{-1}$, see section~\ref{sec:abslines} for more details) due to binary motions. Therefore, one should include the standard error of the average velocity in the model, particularly when we use the absorption velocities. For simplicity, we assume that the uncertainties are uncorrelated and Gaussian distributed. Therefore, we add the primary sources of uncertainty in quadrature to estimate the total velocity uncertainty.  In the next section, we will infer the velocity dispersion of LBVs by including and excluding the standard error of the mean to test whether we are imposing a bias by assuming that the systematic uncertainties are uncorrelated Gaussian distributed.

To find the posterior distribution, we use a two dimensional Monte Carlo Markov chain package (MCMC), \textsc{emcee} \citep{G10,F13} to infer two model parameters ($\mu$ and $\sigma$). For each MCMC run, we use 30 walkers, 2000 steps each, and we burn 1000 of those. For the results presented above, the acceptance fraction is in the $\alpha=0.7$ range.

\subsection{Velocity Dispersion of LBVs vs RSGs}
\label{sec:Disp}
\begin{figure}
\includegraphics[scale=0.5,angle=0]{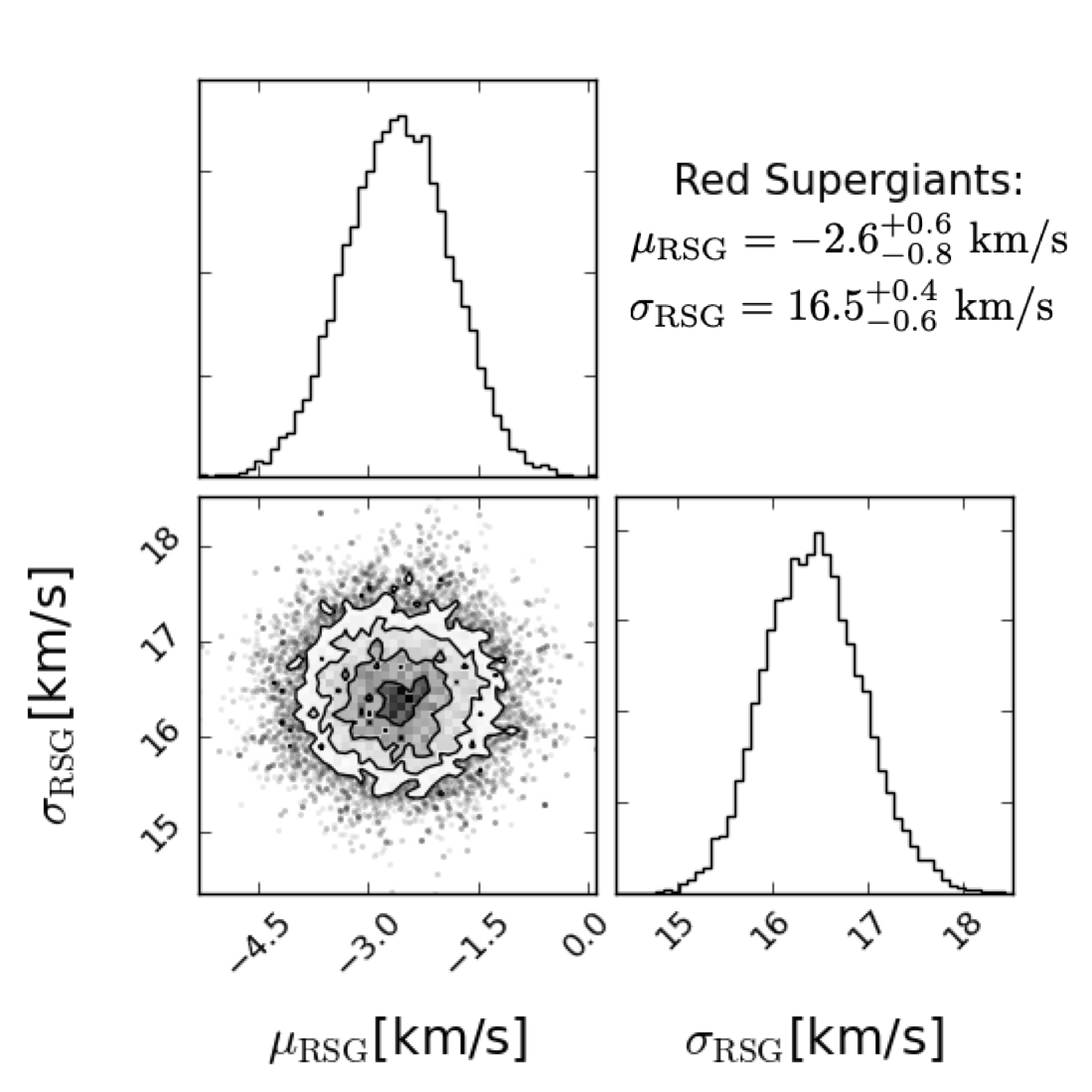}
\caption{The posterior distribution for the two-parameter model. We report the mode and the highest density 68\% confidence interval for the mean ($\mu_{\text{RSG}}$) and the velocity dispersion of RSGs ($\sigma_{\text{RSG}}$). The velocity dispersion of RSGs is 16.5$^{+0.4}_{-0.6}$ km s$^{-1}$, which is approximately 2.5 times smaller than the inferred velocity dispersion of LBVs.}\label{fig:RSGcornerRotationCurve}
\end{figure}

Fig.~\ref{fig:RSGcornerRotationCurve} shows the posterior distribution for $\mu_{\text{RSG}}$ and $\sigma_{\text{RSG}}$.  The values in the top right corner show the mode and the highest 68\% density interval (HDI) for model parameters. The mean velocity of RSG distribution is $\mu_{\text{RSG}}=-2.6^{+0.6}_{-0.8}$ km s$^{-1}$ and velocity dispersion of RSG is $\sigma_{\text{RSG}}=16.5^{+0.4}_{-0.6}$ km s$^{-1}$. This is in good agreement with the simpler estimate in the previous section.   The RSG velocity dispersion is an example of the velocity dispersion of evolved stars. In addition to the velocity dispersion due to drift velocities inherited from their birth cluster, the observed RSG velocity dispersion will also include any dynamical ejections and any kicks imparted by a more massive companion star's previous SN explosion. Therefore, RSG's velocity dispersion is an ideal reference of evolved massive stars to compare to LBVs, in order to determine if LBVs have unusual kinematics that may help explain their anomalous isolation. 

\begin{figure}
\includegraphics[scale=0.5,angle=0]{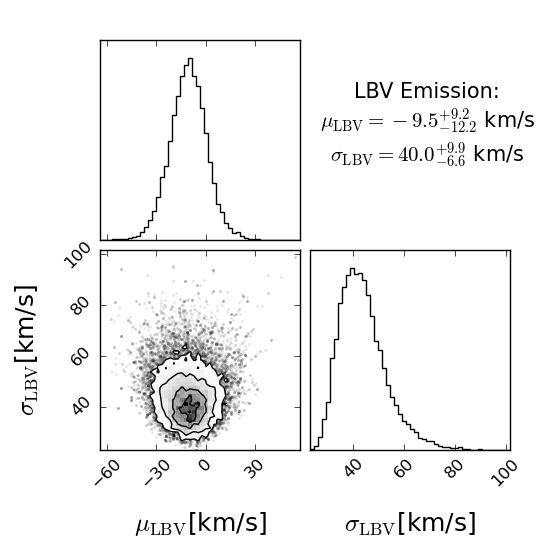}
\caption{The posterior distribution for the mean and velocity dispersion of LBVs. We infer that the velocity dispersion of LBVs  ($\sigma_{\text{LBV}}$) is 40.0$^{+9.9}_{-6.6}$ km s$^{-1}$, which is more than twice the inferred velocity dispersion of RSGs, see Fig.~\ref{fig:RSGcornerRotationCurve}.}\label{fig:LBVcornerRotationCurveEmission}
\end{figure}

Fig.~\ref{fig:LBVcornerRotationCurveEmission} shows the posterior distribution for $\mu_{\text{LBV}}$ and $\sigma_{\text{LBV}}$ using emission lines. The mean velocity of the LBV distribution is $\mu_{\text{LBV}}=-9.5^{+9.2}_{-12.2}$ km s$^{-1}$ and the velocity dispersion of LBVs is $\sigma_{\text{LBV}}=40.0^{+9.9}_{-6.6}$ km s$^{-1}$. Again, this is in good agreement with the simpler estimate above.  We should emphasize that in some cases (S Dor, and R143), emission lines with P Cygni and inverse P Cygni profiles are used to estimate radial velocities and to infer the velocity dispersion of LBVs. However, as we mentioned in section~\ref{sec:velocities}, the absorption parts of these lines are small and we use the emission components of the lines to estimate the radial velocities. As a test, we also remove these targets and we find that the velocity dispersion of LBVs is consistent with the results shown in Figs.~\ref{fig:LBVRotationCurveEmission} and \ref{fig:LBVcornerRotationCurveEmission}.

\begin{figure}
\includegraphics[scale=0.5,angle=0]{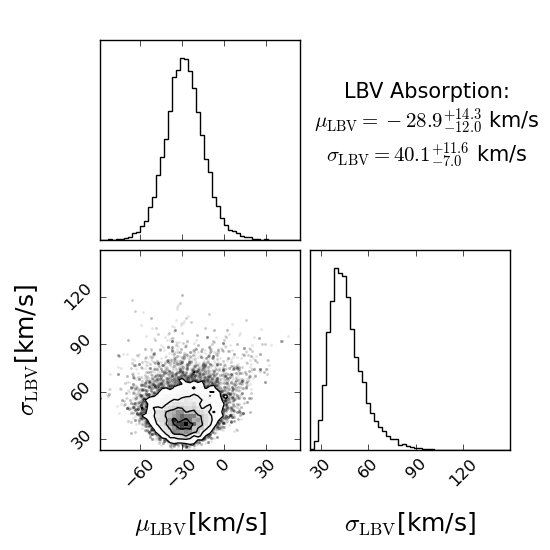}
\caption{Same as Fig.~\ref{fig:LBVcornerRotationCurveEmission}, but using absorption lines. The velocity dispersion of LBVs using absorption lines is consistent with the emission velocity dispersion of LBVs. The absorption velocity distribution is more blueshifted in comparison to emission velocity dispersion, which implies that strong winds may shift the centroid of LBV absorption lines.}\label{fig:LBVcornerRotationCurveAbsorption}
\end{figure}

Similarly, Fig.~\ref{fig:LBVcornerRotationCurveAbsorption} shows the posterior distribution for LBVs using absorption-line radial velocities. The velocity dispersions of LBVs are consistent using either emission and absorption lines, but the centroid of the velocity distribution is more blueshifted using absorption lines due to wind absorption (see section~\ref{sec:centroid} for details).  Because the absorption velocities are more influenced by wind absorption and by binary motion, we take the emission velocities as better indicators of the systemic velocities of LBVs.

We also infer the model parameters for LBVs by excluding the systematic errors due to binary motions mentioned in section~\ref{sec:Bayes}. We find that the mean and velocity dispersion of LBVs using absorption lines are $\mu_{\text{LBV}}=-27.5^{+12.5}_{-13.2}$ and  $\sigma_{\text{LBV}}=40.0^{+13.5}_{-5.7}$ km s$^{-1}$ respectively. The latter results are consistent with the results in Fig.~\ref{fig:LBVcornerRotationCurveAbsorption}, where we include the standard error of the mean velocity in quadrature. Therefore, assuming the systematic errors due to binary motions are uncorrelated and Gaussian distributed does not impose a bias in this sample. Using Bayesian inference, we again find that the velocity dispersion of LBVs is more than twice the inferred velocity dispersion of RSGs. This confirms that the kinematics of LBVs are indeed unusual compared to other evolved massive stars.  

\begin{figure}
\includegraphics[scale=0.5,angle=0]{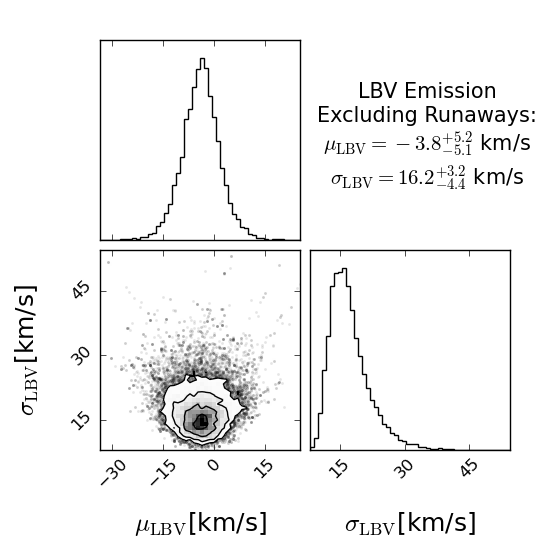}
\caption{Same as Fig.~\ref{fig:LBVcornerRotationCurveEmission}, but excluding R71, S Dor, S119, and Sk-69$^\circ$271, which are the four LBVs with the highest velocities. This is merely an experiment to see what fraction of the LBV sample must be removed to make its inferred velocity dispersion consistent with that of RSGs.}\label{fig:excludingrunaways}
\end{figure}

How much of this large dispersion is attributable to the few fastest runaways?  If we exclude the four fastest LBVs (S119, R71, S Dor, and Sk-69$^\circ$271, which make up 1/4 of the sample), the inferred velocity dispersion of LBVs drops down to about 16 km s$^{-1}$ and becomes consistent with that of RSGs (see Fig.~\ref{fig:excludingrunaways}). We emphasize that excluding the 4 fastest LBVs is merely an experiment to test how much they influence the inferred velocity dispersion of LBVs. There is no legitimate reason to exclude these four LBVs from the velocity dispersion.

\begin{figure}
\includegraphics[scale=0.5,angle=0]{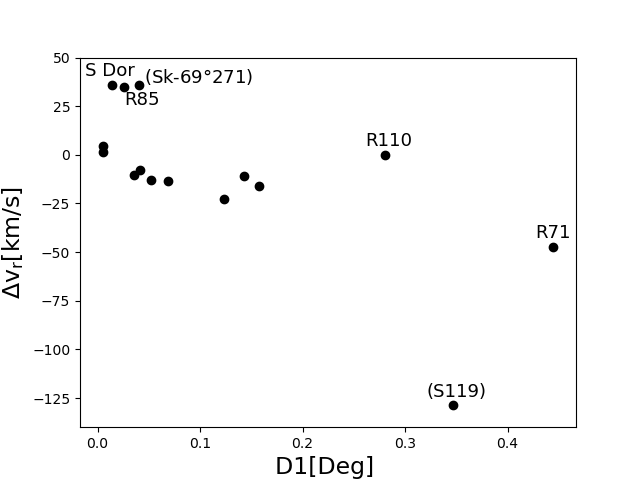}
\caption{$\Delta \rm{v_r}$ values 
calculated using emission lines versus projected distance to the nearest O-type star (D1). D1 values are adopted from \citet{st15}. S119 and R71 are the most isolated ones due to their high velocities, while S Dor, R85, and Sk-69$^\circ$271 are likely slow runaways. The rest of the population are isolated from O-type stars by being rejuvenated either by mass accretion or by being the product of a stellar merger. The extreme isolation of R110 may suggest that R110 is moving sideways, resulting in a low apparent radial velocity.}\label{fig:VelvsSep}
\end{figure}

Fig.~\ref{fig:VelvsSep} shows the relative velocity of LBVs ($\Delta \rm{v_r}$) compared to their projected distances on the sky to the nearest O star (the measured value D1 reported by \citet{st15}). LBVs with the highest velocities, S119 and R71, are the most isolated LBVs from O-type stars. This agreement may hint that the unusual kinematics play a key role in their unusual isolation.  S Dor, R85, and Sk-69$^\circ$271 have lower velocities (25-36 km s$^{-1}$) and are less isolated compared to S119 and R71. The rest of the population have low velocities ($<$25 km s$^{-1}$) but a range of isolation, so it is possible that these appear isolated by being rejuvenated either by mass accretion or by being the product of a stellar merger. Some of them may also have small radial velocities --- but relatively large proper motions --- due to having a trajectory near the plane of the sky as noted earlier; we cannot address this with currently available data.  In particular, R110 is an interesting case with a low radial velocity while being among the most isolated LBVs in the LMC. If R110 has a low projected radial velocity because its direction of motion is near the plane of the sky, it would be a good candidate to examine for measurable proper motion.

\subsection{Centroid Distribution of LBVs vs RSGs}\label{sec:centroid}

The centroid of the LBV distribution is skewed to blueshifted velocities with a larger offset than RSGs. The reason for this offset is unclear, but it could potentially arise for a few reasons: (1) a peculiar spatial distribution of LBVs, (2) absorption in the strong winds of LBVs, or (3) redshifted LBVs that are missing from the sample due to extinction. We discuss these three possibilities below.

\begin{figure}
\includegraphics[width=\linewidth,scale=0.5,angle=0]{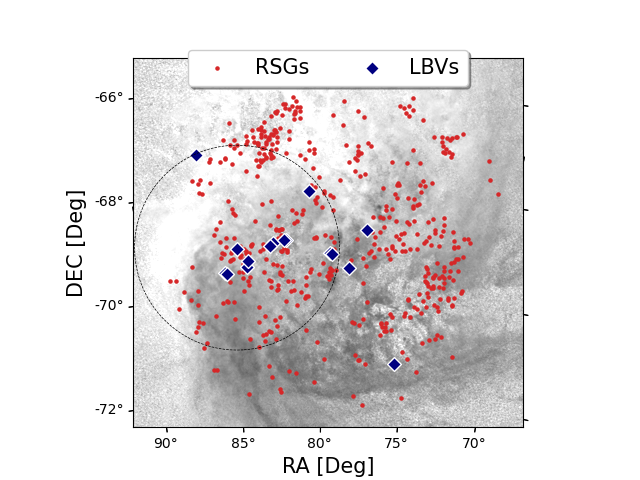}
\caption{Distribution of RSGs (red) and LBVs (blue) projected on the sky as overlaid on the H~{\sc i} gas distribution \citep{K98}. The dashed circle marks a region
that is 2$^\circ$ from the center of 30 Dor. LBVs are concentrated more toward 30 Dor, while the RSGs are spread all over the galaxy. Some redshifted LBVs may be missing due to extinction or other selection biases.}\label{fig:position}
\end{figure}

\begin{figure}
\includegraphics[scale=0.5,angle=0]{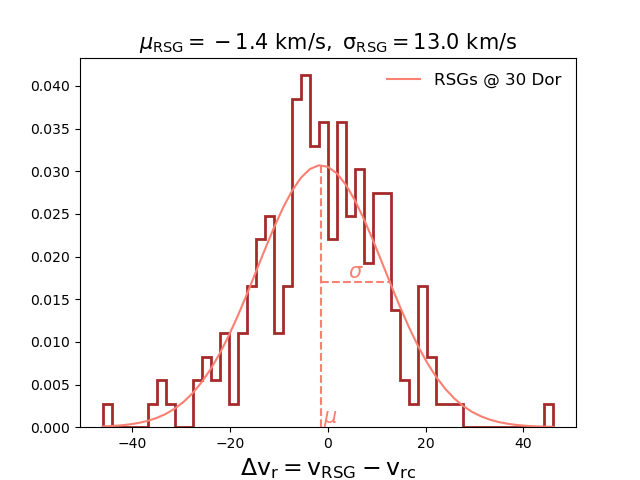}
\caption{Distribution of $\Delta \rm{v_r}$ values for RSGs within
2$^\circ$ of 30 Dor, where LBVs are concentrated. The centroid of the velocity distribution has a similar offset value compared to the full sample of RSGs. This suggests that the stronger blueshift offset observed in LBVs distribution may not be related to the concentration of LBVs near 30 Dor.}\label{fig:SubRSG}
\end{figure}

\begin{figure}
\includegraphics[scale=0.5,angle=0]{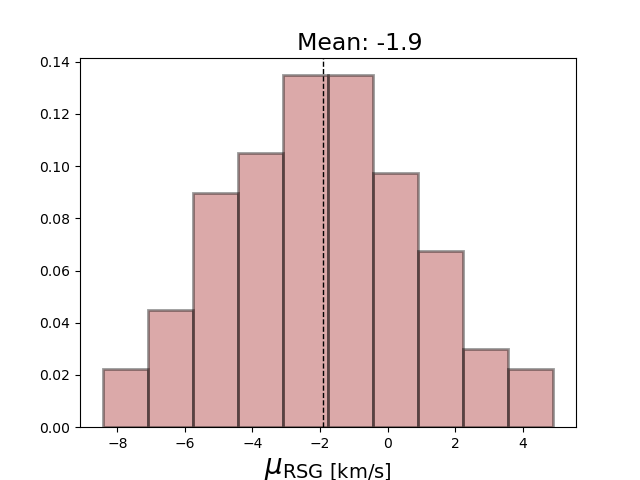}
\caption{Histograms of the centroid velocity for a sample of 16 randomly selected RSGs within 2$^\circ$ of 30 Dor, where the random selection is repeated 100 times. The average centroid velocity is approximately -1.9 km s$^{-1}$ and 97\% of the time the centroid is less blueshifted ($\lvert\mu_{\text{RSG}}\rvert<$ 7 km s$^{-1}$) compared to the centroid velocity of LBVs.}\label{fig:SubRSGHist}
\end{figure}

1) LBVs are somewhat concentrated near 30 Dor, while the RSGs are more evenly distributed across the galaxy (see Fig.~\ref{fig:position}). We expect a net blueshift in the vicinity of 30 Dor due to the clockwise rotation of the galaxy and the inclination of the disk. However, using the rotation curve model of the LMC, we have corrected for the rotation of LMC and we, therefore, subtract out this blueshift. To further investigate whether the larger velocity offset of LBVs arises from peculiar motion arising from their concentration near 30 Dor, we select a sub-sample of RSGs in the same vicinity of 30 Dor as the LBVs. Selecting RSGs within 2 degrees of 30 Dor, we find that the centroid distribution of this subsample of RSGs does not change (see Fig.~\ref{fig:SubRSG}). We also investigate the probability that a net blueshift can be introduced by a combination of a peculiar location and a small sample size.  Instead of taking all the RSGs within 2$^{\circ}$ of 30 Dor, we randomly select 16 RSGs in this region (to match the sample size of LBVs), measure the centroid of that small sample, and then repeat this experiment 100 times. Fig.~\ref{fig:SubRSGHist} shows a histogram of those centroids measured from 100 samples of 16 randomly selected RSGs near 30 Dor. The average centroid distribution of RSGs is approximately -1.9 km s$^{-1}$, and only 3\% of the samples have a centroid comparable to that of the LBVs. These tests indicate that the net blueshift of LBVs is probably not due to their particular location in the LMC or the small sample size.

2) Strong winds may affect the centroids of LBV spectral lines, especially absorption lines formed in the approaching portions of the wind as noted earlier. Figs.~\ref{fig:LBVRotationCurveEmission} and \ref{fig:LBVRotationCurveAbsorption} show that the LBV distribution is more blueshifted using absorption lines in comparison to using emission lines, which is consistent with wind absorption affecting the centroid of absorption lines.  However, if wind absorption were influencing the centroid of emission lines, one would expect it to cause a net redshift in the emission components because of P Cygni absorption. Since we measure a net blueshift even in the emission lines, and this blueshift is less than in absorption lines, it is unlikely that this blueshift arises from wind effects.
 
3) Some LBVs that are redshifted may be missing due to foreground extinction. Due to an excess of dense molecular gas in the vicinity of 30 Dor, redshifted LBVs located on the far side of the clouds may look less luminous and less massive than one would expect for massive LBVs. Stars are usually classified as LBVs based on their decade-long visible variability, combined with a high inferred luminosity. If redshifted LBVs on the far side of 30 Dor suffer a couple of magnitudes of extinction in the visual band, they may be interpreted as lower-mass stars, and therefore have not been considered as LBVs.  RSGs are less affected by foreground reddening (many are selected based on their red colors and IR excess), and also they are relatively older stars that have drifted farther away from their natal dust clouds in comparison to LBVs.  This hypothesis seems plausible but difficult to prove without more investigation; such reddened LBVs might have been interpreted as reddened B supergiants and might not have been as closely monitored for variability.

\begin{figure}
\includegraphics[scale=0.5,angle=0]{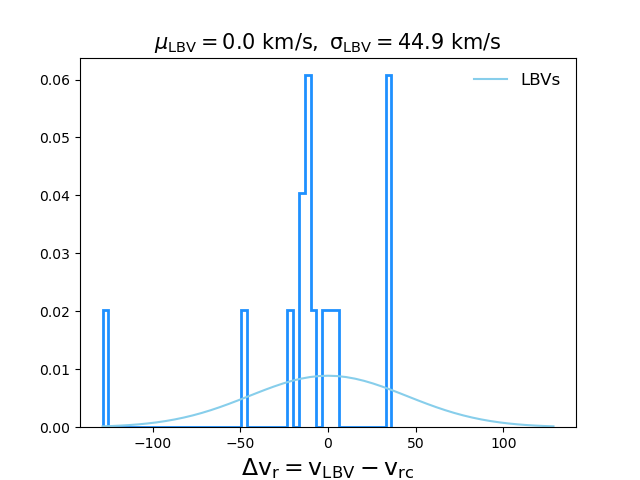}
\caption{Same as Fig.~\ref{fig:LBVRotationCurveEmission}, but the Gaussian curve is a fit to only the blue side of the distribution, assuming a symmetric distribution and the centroid of the distribution is forced to be at zero. If some LBVs are missing due to extinction or other selection biases, the velocity dispersion of LBVs should be larger than the current estimated value.  }\label{fig:ShiftedHist}
\end{figure}

\begin{figure}
\includegraphics[scale=0.5,angle=0]{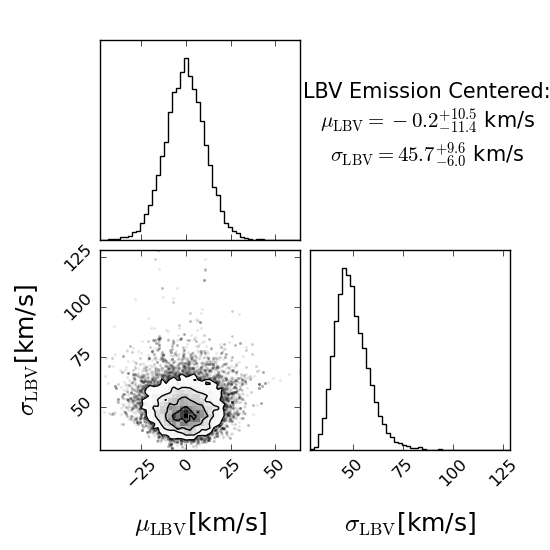}
\caption{Same as Fig.~\ref{fig:LBVcornerRotationCurveEmission}, but the centroid of the distribution is forced to be around zero. The Gaussian distribution is shown in Fig.~\ref{fig:ShiftedHist}.}\label{fig:LBVshifted}
\end{figure}

If the net blueshift is caused by a selection effect that excludes some redshifted LBVs, then the true velocity dispersion of LBVs may be even larger than what we inferred. Figs.~\ref{fig:ShiftedHist} and \ref{fig:LBVshifted} show the histogram and posterior distribution for LBVs, where this time we force the centroid of the distribution to be at 0 km s$^{-1}$, and we fit only the blueshifted side of the distribution.  Using this method, we infer a higher velocity dispersion of $\sigma_{\text{LBV}}=45.7^{+9.6}_{-6.0}$ km s$^{-1}$.  This would not, however, change the conclusions we present here.  The main point of our investigation is that LBVs have a larger velocity dispersion than RSGs, even if we ignore any possible missing redshifted LBVs (which would raise that dispersion value). As noted above, this is more than twice the value found for RSGs, indicating that LBVs contain a larger fraction of runaways than most other evolved massive stars.

\section{Conclusion}\label{sec:consclusion}
We obtained high-resolution echelle spectra of 16 LBVs in the LMC using the MIKE spectrograph on the Magellan Clay Telescope. We estimate the average heliocentric radial velocity of LMC LBVs using wind emission, nebular emission lines (when available), and absorption lines.  On average, the radial velocities measured in this work are in good agreement with the previously published radial velocities in the literature.  The absorption velocities in most cases are blueshifted compared to emission velocities, which indicates that they are affected by wind absorption. We also find that a few of our targets have variable absorption velocities over time due to the Doppler shifts from orbital motion in binaries.  This orbital motion appears to have little influence on the emission-line velocities, so we take emission lines to best represent their systemic velocities.  

In the traditional single-star scenario, LBVs should be found in or near young clusters along with other O-type stars, whereas observations reveal otherwise \citep{st15}. LBVs appear isolated from main-sequence massive stars either because they received a significant kick when their companion in a binary system exploded, or because their true lifetime is longer than we expect based on their current luminosity --- i.e. they were rejuvenated either by mass accretion in a binary (without a significant kick from a companion's SN) or by being the product of a stellar merger \citep{st15,mojgan17}. In this paper, we investigate to what degree LBVs have unusual kinematics that might help cause their anomalous isolation.  To constrain whether LMC LBVs have peculiar velocities, we compare the observed radial velocity of each LBV with the projected rotation curve velocity \citep{C22} at LBV's location to estimate how fast they move with respect to their local environment. 

In order to determine if LBVs have peculiar kinematics compared to other evolved massive stars, we compare the LBV velocity dispersion with that of RSGs. Using Bayesian inference, we find that the dispersion of LBVs is 40.0$^{+9.9}_{-6.6}$ km s$^{-1}$, which is more than twice the inferred velocity dispersion of RSGs. If we exclude the four high-velocity LBVs (S119, R71, S Dor, and Sk-69$^\circ$271) from the inference, the velocity dispersion of LBVs will reduce to
16.2$^{+3.2}_{-4.4}$ km s$^{-1}$, which is consistent with the velocity dispersion of RSGs. This test measures how much high-velocity LBVs affect the inferred velocity dispersion but there is no physical reason to exclude these four stars from LBVs’ population. High-velocity LBVs indicate the typical motions of the population and suggest that, on average, LBVs should exhibit high tangential velocity. Due to {\it Gaia} EDR3 negative parallaxes and large uncertainties, currently, we are unable to measure LBVs in-plane velocities. The situation may improve in the future with {\it Gaia} DR4.

Taking $\lvert\Delta \rm{v_r} \rvert\ge$ 25 km s$^{-1}$ as a relatively arbitrary way to designate stars with high velocities compared to the local rotation curve, we find that 33\% of LBVs have such high velocities, whereas only 9\% of RSGs show these fast motions.  This detected fraction of 33\% is a lower limit to the true fraction of LBVs that are runaways, since some fast-moving LBVs may have low Doppler shifts if their motion is near the plane of the sky (as may be the case for R110).  We therefore find it likely that about half of LBVs are relatively fast runaways/walkaways.

The remaining half of LBVs that are slow-moving may still have received a kick from a companion's SN, since such kicks are expected to be small \citep{Renzo}. Alternatively, these slow LBVs might be massive blue stragglers that are rejuvenated either by mass accretion or mergers.  If an LBV is a rejuvenated star, the high-luminosity star we currently see has originated from an initially lower-mass star, allowing the star to be much older than we would infer from its current luminosity.  This longer lifetime allows it to disperse to larger distances even with a small velocity, or provides extra time for its O-type neighbors to die, thereby producing large separations from O-type stars observed by \citet{st15}. In this picture, LBV birth clusters could be relatively nearby but probably dispersed and lacking early O-type stars. Therefore, for LBVs that do not have peculiar velocities relative to the rotation curve model, it will be interesting to study nearby stellar populations to infer their ages.

\section*{Acknowledgements}
We thank Jeremiah W. Murphy and Jose Groh for helpful and inspiring discussions during the preparation of this manuscript. This paper includes data gathered with the 6.5 m Magellan Telescopes located at Las Campanas Observatory, Chile. This work made use of the ASAS-SN Sky Patrol light curve \citep{S14, K17} and the AAVSO International Database.  Support was provided by the National Aeronautics and Space Administration (NASA) through HST grant AR-14316 from the Space Telescope Science Institute, which is operated by AURA, Inc., under NASA contract NAS5-26555.

\section*{Data Availability}
The data underlying this article will be shared on reasonable request to the corresponding author.
\bibliographystyle{mnras}
\DeclareRobustCommand{\VAN}[3]{#3}
\bibliography{MA}

\begin{thebibliography}{}
\makeatletter
\relax
\def\mn@urlcharsother{\let\do\@makeother \do\$\do\&\do\#\do\^\do\_\do\%\do\~}
\def\mn@doi{\begingroup\mn@urlcharsother \@ifnextchar [ {\mn@doi@}
  {\mn@doi@[]}}
\def\mn@doi@[#1]#2{\def\@tempa{#1}\ifx\@tempa\@empty \href
  {http://dx.doi.org/#2} {doi:#2}\else \href {http://dx.doi.org/#2} {#1}\fi
  \endgroup}
\def\mn@eprint#1#2{\mn@eprint@#1:#2::\@nil}
\def\mn@eprint@arXiv#1{\href {http://arxiv.org/abs/#1} {{\tt arXiv:#1}}}
\def\mn@eprint@dblp#1{\href {http://dblp.uni-trier.de/rec/bibtex/#1.xml}
  {dblp:#1}}
\def\mn@eprint@#1:#2:#3:#4\@nil{\def\@tempa {#1}\def\@tempb {#2}\def\@tempc
  {#3}\ifx \@tempc \@empty \let \@tempc \@tempb \let \@tempb \@tempa \fi \ifx
  \@tempb \@empty \def\@tempb {arXiv}\fi \@ifundefined
  {mn@eprint@\@tempb}{\@tempb:\@tempc}{\expandafter \expandafter \csname
  mn@eprint@\@tempb\endcsname \expandafter{\@tempc}}}

\bibitem[\protect\citeauthoryear{{Aadland}, {Massey}, {Neugent}  \&
  {Drout}}{{Aadland} et~al.}{2018}]{Aadland18}
{Aadland} E.,  {Massey} P.,  {Neugent} K.~F.,   {Drout} M.~R.,  2018, \mn@doi
  [\aj] {10.3847/1538-3881/aaeb96}, \href
  {https://ui.adsabs.harvard.edu/abs/2018AJ....156..294A} {156, 294}

\bibitem[\protect\citeauthoryear{{Aghakhanloo} et~al.}{{Aghakhanloo}
  et~al.}{2017}]{mojgan17}
{Aghakhanloo} M.,  et~al., 2017, Monthly Notices of the Royal Astronomical
  Society, \href {https://ui.adsabs.harvard.edu/abs/2017MNRAS.472..591A} {472,
  591}

\bibitem[\protect\citeauthoryear{{Agliozzo}, {Umana}, {Trigilio}, {Buemi},
  {Leto}, {Ingallinera}, {Franzen}  \& {Noriega-Crespo}}{{Agliozzo}
  et~al.}{2012}]{A12}
{Agliozzo} C.,  {Umana} G.,  {Trigilio} C.,  {Buemi} C.,  {Leto} P.,
  {Ingallinera} A.,  {Franzen} T.,   {Noriega-Crespo} A.,  2012, \mn@doi
  [\mnras] {10.1111/j.1365-2966.2012.21791.x}, \href
  {https://ui.adsabs.harvard.edu/abs/2012MNRAS.426..181A} {426, 181}

\bibitem[\protect\citeauthoryear{Aldoretta et~al.,}{Aldoretta
  et~al.}{2014}]{A14}
Aldoretta E.~J.,  et~al., 2014, \mn@doi [The Astronomical Journal]
  {10.1088/0004-6256/149/1/26}, 149, 26

\bibitem[\protect\citeauthoryear{{Ardeberg}, {Brunet}, {Maurice}  \&
  {Prevot}}{{Ardeberg} et~al.}{1972}]{A72}
{Ardeberg} A.,  {Brunet} J.~P.,  {Maurice} E.,   {Prevot} L.,  1972, \aaps,
  \href {https://ui.adsabs.harvard.edu/abs/1972A&AS....6..249A} {6, 249}

\bibitem[\protect\citeauthoryear{Bernstein et~al.}{Bernstein
  et~al.}{2003}]{B03}
Bernstein R.,  et~al., 2003, in Instrument Design and Performance for
  Optical/Infrared Ground-based Telescopes. pp 1694--1704

\bibitem[\protect\citeauthoryear{{Besla}, {Kallivayalil}, {Hernquist}, {van der
  Marel}, {Cox}  \& {Kere{\v{s}}}}{{Besla} et~al.}{2012}]{B12}
{Besla} G.,  {Kallivayalil} N.,  {Hernquist} L.,  {van der Marel} R.~P.,  {Cox}
  T.~J.,   {Kere{\v{s}}} D.,  2012, \mn@doi [\mnras]
  {10.1111/j.1365-2966.2012.20466.x}, \href
  {https://ui.adsabs.harvard.edu/abs/2012MNRAS.421.2109B} {421, 2109}

\bibitem[\protect\citeauthoryear{{Besla}, {Mart{\'\i}nez-Delgado}, {van der
  Marel}, {Beletsky}, {Seibert}, {Schlafly}, {Grebel}  \& {Neyer}}{{Besla}
  et~al.}{2016}]{B16}
{Besla} G.,  {Mart{\'\i}nez-Delgado} D.,  {van der Marel} R.~P.,  {Beletsky}
  Y.,  {Seibert} M.,  {Schlafly} E.~F.,  {Grebel} E.~K.,   {Neyer} F.,  2016,
  \mn@doi [\apj] {10.3847/0004-637X/825/1/20}, \href
  {https://ui.adsabs.harvard.edu/abs/2016ApJ...825...20B} {825, 20}

\bibitem[\protect\citeauthoryear{{Blaauw}}{{Blaauw}}{1961}]{B61}
{Blaauw} A.,  1961, \bain, \href
  {https://ui.adsabs.harvard.edu/abs/1961BAN....15..265B} {15, 265}

\bibitem[\protect\citeauthoryear{{Campagnolo}, {Borges Fernandes}, {Drake},
  {Kraus}, {Guerrero}  \& {Pereira}}{{Campagnolo} et~al.}{2018}]{C18}
{Campagnolo} J.~C.~N.,  {Borges Fernandes} M.,  {Drake} N.~A.,  {Kraus} M.,
  {Guerrero} C.~A.,   {Pereira} C.~B.,  2018, \mn@doi [\aap]
  {10.1051/0004-6361/201731785}, \href
  {https://ui.adsabs.harvard.edu/abs/2018A&A...613A..33C} {613, A33}

\bibitem[\protect\citeauthoryear{{Choi}, {Olsen}, {Besla}, {van der Marel},
  {Zivick}, {Kallivayalil}  \& {Nidever}}{{Choi} et~al.}{2022}]{C22}
{Choi} Y.,  {Olsen} K. A.~G.,  {Besla} G.,  {van der Marel} R.~P.,  {Zivick}
  P.,  {Kallivayalil} N.,   {Nidever} D.~L.,  2022, arXiv e-prints, \href
  {https://ui.adsabs.harvard.edu/abs/2022arXiv220104648C} {p. arXiv:2201.04648}

\bibitem[\protect\citeauthoryear{Collaboration, Brown, Vallenari, Prusti, de
  Bruijne, Babusiaux  \& Biermann}{Collaboration et~al.}{2020}]{g20}
Collaboration G.,  Brown A. G.~A.,  Vallenari A.,  Prusti T.,  de Bruijne J.
  H.~J.,  Babusiaux C.,   Biermann M.,  2020, Gaia Early Data Release 3:
  Summary of the contents and survey properties (\mn@eprint {arXiv}
  {2012.01533})

\bibitem[\protect\citeauthoryear{{Conti}}{{Conti}}{1975}]{C75}
{Conti} P.~S.,  1975, Memoires of the Societe Royale des Sciences de Liege,
  \href {https://ui.adsabs.harvard.edu/abs/1975MSRSL...9..193C} {9, 193}

\bibitem[\protect\citeauthoryear{{Conti}}{{Conti}}{1984}]{C84}
{Conti} P.~S.,  1984, in {Maeder} A.,  {Renzini} A.,  eds,  IAU Symposium Vol.
  105, Observational Tests of the Stellar Evolution Theory. p.~233

\bibitem[\protect\citeauthoryear{{Danforth} \& {Chu}}{{Danforth} \&
  {Chu}}{2001}]{D01}
{Danforth} C.~W.,  {Chu} Y.-H.,  2001, \mn@doi [\apjl] {10.1086/320341}, \href
  {https://ui.adsabs.harvard.edu/abs/2001ApJ...552L.155D} {552, L155}

\bibitem[\protect\citeauthoryear{{Evans}, {van Loon}, {Hainich}  \&
  {Bailey}}{{Evans} et~al.}{2015}]{E15}
{Evans} C.~J.,  {van Loon} J.~T.,  {Hainich} R.,   {Bailey} M.,  2015, \mn@doi
  [\aap] {10.1051/0004-6361/201525882}, \href
  {https://ui.adsabs.harvard.edu/abs/2015A&A...584A...5E} {584, A5}

\bibitem[\protect\citeauthoryear{{Feast}, {Thackeray}  \& {Wesselink}}{{Feast}
  et~al.}{1960}]{F60}
{Feast} M.~W.,  {Thackeray} A.~D.,   {Wesselink} A.~J.,  1960, \mn@doi [\mnras]
  {10.1093/mnras/121.4.337}, \href
  {https://ui.adsabs.harvard.edu/abs/1960MNRAS.121..337F} {121, 337}

\bibitem[\protect\citeauthoryear{Foreman-Mackey, Hogg, Lang  \&
  Goodman}{Foreman-Mackey et~al.}{2013}]{F13}
Foreman-Mackey D.,  Hogg D.~W.,  Lang D.,   Goodman J.,  2013, \mn@doi [PASP]
  {10.1086/670067}, 125, 306

\bibitem[\protect\citeauthoryear{{Gaia Collaboration} et~al.,}{{Gaia
  Collaboration} et~al.}{2021}]{GaiaEDR3}
{Gaia Collaboration} et~al., 2021, \mn@doi [\aap]
  {10.1051/0004-6361/202039657}, \href
  {https://ui.adsabs.harvard.edu/abs/2021A&A...649A...1G} {649, A1}

\bibitem[\protect\citeauthoryear{Goodman \& Weare}{Goodman \&
  Weare}{2010}]{G10}
Goodman J.,  Weare J.,  2010, \mn@doi [Commun. Appl. Math. Comput. Sci.]
  {10.2140/camcos.2010.5.65}, 5, 65

\bibitem[\protect\citeauthoryear{{Groh} et~al.,}{{Groh} et~al.}{2009}]{G09}
{Groh} J.~H.,  et~al., 2009, \mn@doi [\apjl] {10.1088/0004-637X/705/1/L25},
  \href {https://ui.adsabs.harvard.edu/abs/2009ApJ...705L..25G} {705, L25}

\bibitem[\protect\citeauthoryear{{Groh}, {Hillier}  \& {Damineli}}{{Groh}
  et~al.}{2011}]{G11}
{Groh} J.~H.,  {Hillier} D.~J.,   {Damineli} A.,  2011, \mn@doi [\apj]
  {10.1088/0004-637X/736/1/46}, \href
  {https://ui.adsabs.harvard.edu/abs/2011ApJ...736...46G} {736, 46}

\bibitem[\protect\citeauthoryear{{Hubble} \& {Sandage}}{{Hubble} \&
  {Sandage}}{1953}]{HS53}
{Hubble} E.,  {Sandage} A.,  1953, \mn@doi [\apj] {10.1086/145764}, \href
  {http://adsabs.harvard.edu/abs/1953ApJ...118..353H} {118, 353}

\bibitem[\protect\citeauthoryear{{Humphreys} \& {Davidson}}{{Humphreys} \&
  {Davidson}}{1994}]{hd94}
{Humphreys} R.~M.,  {Davidson} K.,  1994, \mn@doi [\pasp] {10.1086/133478},
  \href {https://ui.adsabs.harvard.edu/abs/1994PASP..106.1025H} {106, 1025}

\bibitem[\protect\citeauthoryear{{Humphreys}, {Weis}, {Davidson}  \&
  {Gordon}}{{Humphreys} et~al.}{2016}]{H16}
{Humphreys} R.~M.,  {Weis} K.,  {Davidson} K.,   {Gordon} M.~S.,  2016, \mn@doi
  [\apj] {10.3847/0004-637X/825/1/64}, \href
  {https://ui.adsabs.harvard.edu/abs/2016ApJ...825...64H} {825, 64}

\bibitem[\protect\citeauthoryear{{Humphreys}, {Davidson}, {Hahn}, {Martin}  \&
  {Weis}}{{Humphreys} et~al.}{2017}]{H17}
{Humphreys} R.~M.,  {Davidson} K.,  {Hahn} D.,  {Martin} J.~C.,   {Weis} K.,
  2017, \mn@doi [\apj] {10.3847/1538-4357/aa7cef}, \href
  {https://ui.adsabs.harvard.edu/abs/2017ApJ...844...40H} {844, 40}

\bibitem[\protect\citeauthoryear{{Justham}, {Podsiadlowski}  \&
  {Vink}}{{Justham} et~al.}{2014}]{J14}
{Justham} S.,  {Podsiadlowski} P.,   {Vink} J.~S.,  2014, \mn@doi [\apj]
  {10.1088/0004-637X/796/2/121}, \href
  {https://ui.adsabs.harvard.edu/abs/2014ApJ...796..121J} {796, 121}

\bibitem[\protect\citeauthoryear{Kim et~al.}{Kim et~al.}{1998}]{K98}
Kim S.,  et~al., 1998, The Astrophysical Journal, 503, 674

\bibitem[\protect\citeauthoryear{{Kochanek} et~al.,}{{Kochanek}
  et~al.}{2017}]{K17}
{Kochanek} C.~S.,  et~al., 2017, \mn@doi [\pasp] {10.1088/1538-3873/aa80d9},
  \href {https://ui.adsabs.harvard.edu/abs/2017PASP..129j4502K} {129, 104502}

\bibitem[\protect\citeauthoryear{{Kramida}, {Ralchenko}  \& {Reader}}{{Kramida}
  et~al.}{2016}]{K16}
{Kramida} A.,  {Ralchenko} Y.,   {Reader} J.,  2016, in APS Division of Atomic,
  Molecular and Optical Physics Meeting Abstracts. p. Q1.202

\bibitem[\protect\citeauthoryear{{Kraus}}{{Kraus}}{2019}]{K19}
{Kraus} M.,  2019, \mn@doi [Galaxies] {10.3390/galaxies7040083}, \href
  {https://ui.adsabs.harvard.edu/abs/2019Galax...7...83K} {7, 83}

\bibitem[\protect\citeauthoryear{{Leonard} \& {Duncan}}{{Leonard} \&
  {Duncan}}{1988}]{L88}
{Leonard} P. J.~T.,  {Duncan} M.~J.,  1988, \mn@doi [\aj] {10.1086/114804},
  \href {https://ui.adsabs.harvard.edu/abs/1988AJ.....96..222L} {96, 222}

\bibitem[\protect\citeauthoryear{{Lindegren} et~al.,}{{Lindegren}
  et~al.}{2018}]{L18}
{Lindegren} L.,  et~al., 2018, \mn@doi [\aap] {10.1051/0004-6361/201832727},
  \href {https://ui.adsabs.harvard.edu/abs/2018A&A...616A...2L} {616, A2}

\bibitem[\protect\citeauthoryear{{Mahy} et~al.,}{{Mahy}
  et~al.}{2022}]{mahy2022}
{Mahy} L.,  et~al., 2022, \mn@doi [\aap] {10.1051/0004-6361/202040062}, \href
  {https://ui.adsabs.harvard.edu/abs/2022A&A...657A...4M} {657, A4}

\bibitem[\protect\citeauthoryear{{Moe} et~al.}{{Moe} et~al.}{2017}]{M17}
{Moe} M.,  et~al., 2017, \mn@doi [The Astrophysical Journal Supplement Series]
  {10.3847/1538-4365/aa6fb6}, \href
  {https://ui.adsabs.harvard.edu/abs/2017ApJS..230...15M} {230, 15}

\bibitem[\protect\citeauthoryear{{Munari} et~al.,}{{Munari} et~al.}{2009}]{M09}
{Munari} U.,  et~al., 2009, \mn@doi [A\&A] {10.1051/0004-6361/200912398}, 503,
  511

\bibitem[\protect\citeauthoryear{{Neugent} et~al.}{{Neugent}
  et~al.}{2012}]{N12}
{Neugent} K.~F.,  et~al., 2012, The Astrophysical Journal, 749, 177

\bibitem[\protect\citeauthoryear{{Nota}, {Pasquali}, {Drissen}, {Leitherer},
  {Robert}, {Moffat}  \& {Schmutz}}{{Nota} et~al.}{1996}]{N96}
{Nota} A.,  {Pasquali} A.,  {Drissen} L.,  {Leitherer} C.,  {Robert} C.,
  {Moffat} A. F.~J.,   {Schmutz} W.,  1996, \mn@doi [\apjs] {10.1086/192263},
  \href {https://ui.adsabs.harvard.edu/abs/1996ApJS..102..383N} {102, 383}

\bibitem[\protect\citeauthoryear{{Oh} \& {Kroupa}}{{Oh} \&
  {Kroupa}}{2016}]{O16}
{Oh} S.,  {Kroupa} P.,  2016, \mn@doi [\aap] {10.1051/0004-6361/201628233},
  \href {https://ui.adsabs.harvard.edu/abs/2016A&A...590A.107O} {590, A107}

\bibitem[\protect\citeauthoryear{{Olsen}, {Zaritsky}, {Blum}, {Boyer}  \&
  {Gordon}}{{Olsen} et~al.}{2011}]{O11}
{Olsen} K. A.~G.,  {Zaritsky} D.,  {Blum} R.~D.,  {Boyer} M.~L.,   {Gordon}
  K.~D.,  2011, \mn@doi [\apj] {10.1088/0004-637X/737/1/29}, \href
  {https://ui.adsabs.harvard.edu/abs/2011ApJ...737...29O} {737, 29}

\bibitem[\protect\citeauthoryear{{Perets} \& {{\v{S}}ubr}}{{Perets} \&
  {{\v{S}}ubr}}{2012}]{P12}
{Perets} H.~B.,  {{\v{S}}ubr} L.,  2012, \mn@doi [\apj]
  {10.1088/0004-637X/751/2/133}, \href
  {https://ui.adsabs.harvard.edu/abs/2012ApJ...751..133P} {751, 133}

\bibitem[\protect\citeauthoryear{{Poveda}, {Ruiz}  \& {Allen}}{{Poveda}
  et~al.}{1967}]{P67}
{Poveda} A.,  {Ruiz} J.,   {Allen} C.,  1967, Boletin de los Observatorios
  Tonantzintla y Tacubaya, \href
  {https://ui.adsabs.harvard.edu/abs/1967BOTT....4...86P} {4, 86}

\bibitem[\protect\citeauthoryear{{Renzo} et~al.}{{Renzo} et~al.}{2019}]{Renzo}
{Renzo} M.,  et~al., 2019, Astronomy \& Astrophysics, 624, A66

\bibitem[\protect\citeauthoryear{{Sana} et~al.}{{Sana} et~al.}{2012}]{S12}
{Sana} H.,  et~al., 2012, Science, 337, 444

\bibitem[\protect\citeauthoryear{{Shappee} et~al.,}{{Shappee}
  et~al.}{2014}]{S14}
{Shappee} B.~J.,  et~al., 2014, \mn@doi [\apj] {10.1088/0004-637X/788/1/48},
  \href {https://ui.adsabs.harvard.edu/abs/2014ApJ...788...48S} {788, 48}

\bibitem[\protect\citeauthoryear{{Smith}}{{Smith}}{2011}]{SN11}
{Smith} N.,  2011, \mn@doi [\mnras] {10.1111/j.1365-2966.2011.18607.x}, \href
  {https://ui.adsabs.harvard.edu/abs/2011MNRAS.415.2020S} {415, 2020}

\bibitem[\protect\citeauthoryear{{Smith}}{{Smith}}{2016}]{S16}
{Smith} N.,  2016, \mn@doi [\mnras] {10.1093/mnras/stw1533}, \href
  {https://ui.adsabs.harvard.edu/abs/2016MNRAS.461.3353S} {461, 3353}

\bibitem[\protect\citeauthoryear{{Smith}}{{Smith}}{2017}]{S17}
{Smith} N.,  2017, \mn@doi [Philosophical Transactions of the Royal Society of
  London Series A] {10.1098/rsta.2016.0268}, \href
  {https://ui.adsabs.harvard.edu/abs/2017RSPTA.37560268S} {375, 20160268}

\bibitem[\protect\citeauthoryear{{Smith}}{{Smith}}{2019}]{S19bs}
{Smith} N.,  2019, \mn@doi [\mnras] {10.1093/mnras/stz2277}, \href
  {https://ui.adsabs.harvard.edu/abs/2019MNRAS.489.4378S} {489, 4378}

\bibitem[\protect\citeauthoryear{{Smith} \& {Tombleson}}{{Smith} \&
  {Tombleson}}{2015}]{st15}
{Smith} N.,  {Tombleson} R.,  2015, \mn@doi [\mnras] {10.1093/mnras/stu2430},
  \href {https://ui.adsabs.harvard.edu/abs/2015MNRAS.447..598S} {447, 598}

\bibitem[\protect\citeauthoryear{{Smith}, {Li}, {Silverman}, {Ganeshalingam}
  \& {Filippenko}}{{Smith} et~al.}{2011}]{S11}
{Smith} N.,  {Li} W.,  {Silverman} J.~M.,  {Ganeshalingam} M.,   {Filippenko}
  A.~V.,  2011, \mn@doi [\mnras] {10.1111/j.1365-2966.2011.18763.x}, \href
  {https://ui.adsabs.harvard.edu/abs/2011MNRAS.415..773S} {415, 773}

\bibitem[\protect\citeauthoryear{{Smith} et~al.,}{{Smith}
  et~al.}{2016a}]{smith2016merger}
{Smith} N.,  et~al., 2016a, \mn@doi [\mnras] {10.1093/mnras/stw219}, \href
  {https://ui.adsabs.harvard.edu/abs/2016MNRAS.458..950S} {458, 950}

\bibitem[\protect\citeauthoryear{{Smith}, {Andrews}  \& {Mauerhan}}{{Smith}
  et~al.}{2016b}]{smith2016sn09ip}
{Smith} N.,  {Andrews} J.~E.,   {Mauerhan} J.~C.,  2016b, \mn@doi [\mnras]
  {10.1093/mnras/stw2190}, \href
  {https://ui.adsabs.harvard.edu/abs/2016MNRAS.463.2904S} {463, 2904}

\bibitem[\protect\citeauthoryear{{Smith} et~al.,}{{Smith}
  et~al.}{2018}]{smith2018}
{Smith} N.,  et~al., 2018, \mn@doi [\mnras] {10.1093/mnras/sty1500}, \href
  {https://ui.adsabs.harvard.edu/abs/2018MNRAS.480.1466S} {480, 1466}

\bibitem[\protect\citeauthoryear{{Smith}, {Aghakhanloo}, {Murphy}, {Drout},
  {Stassun}  \& {Groh}}{{Smith} et~al.}{2019}]{S19}
{Smith} N.,  {Aghakhanloo} M.,  {Murphy} J.~W.,  {Drout} M.~R.,  {Stassun}
  K.~G.,   {Groh} J.~H.,  2019, \mn@doi [\mnras] {10.1093/mnras/stz1712}, \href
  {https://ui.adsabs.harvard.edu/abs/2019MNRAS.488.1760S} {488, 1760}

\bibitem[\protect\citeauthoryear{{Smith} et~al.,}{{Smith} et~al.}{2020}]{s20}
{Smith} N.,  et~al., 2020, \mn@doi [\mnras] {10.1093/mnras/staa061}, \href
  {https://ui.adsabs.harvard.edu/abs/2020MNRAS.492.5897S} {492, 5897}

\bibitem[\protect\citeauthoryear{{Stahl}}{{Stahl}}{1986}]{SO86}
{Stahl} O.,  1986, \aap, \href
  {https://ui.adsabs.harvard.edu/abs/1986A&A...164..321S} {164, 321}

\bibitem[\protect\citeauthoryear{{Stahl}}{{Stahl}}{1987}]{Stahl87}
{Stahl} O.,  1987, \aap, \href
  {https://ui.adsabs.harvard.edu/abs/1987A&A...182..229S} {182, 229}

\bibitem[\protect\citeauthoryear{{Stahl}, {Wolf}  \& {Zickgraf}}{{Stahl}
  et~al.}{1987}]{S87}
{Stahl} O.,  {Wolf} B.,   {Zickgraf} F.~J.,  1987, \aap, \href
  {https://ui.adsabs.harvard.edu/abs/1987A&A...184..193S} {184, 193}

\bibitem[\protect\citeauthoryear{{Staveley-Smith}, {Kim}, {Calabretta},
  {Haynes}  \& {Kesteven}}{{Staveley-Smith} et~al.}{2003}]{S03}
{Staveley-Smith} L.,  {Kim} S.,  {Calabretta} M.~R.,  {Haynes} R.~F.,
  {Kesteven} M.~J.,  2003, \mn@doi [\mnras] {10.1046/j.1365-8711.2003.06146.x},
  \href {https://ui.adsabs.harvard.edu/abs/2003MNRAS.339...87S} {339, 87}

\bibitem[\protect\citeauthoryear{{Tammann} \& {Sandage}}{{Tammann} \&
  {Sandage}}{1968}]{T68}
{Tammann} G.~A.,  {Sandage} A.,  1968, \mn@doi [\apj] {10.1086/149487}, \href
  {https://ui.adsabs.harvard.edu/abs/1968ApJ...151..825T} {151, 825}

\bibitem[\protect\citeauthoryear{{Tubbesing} et~al.,}{{Tubbesing}
  et~al.}{2002}]{T02}
{Tubbesing} S.,  et~al., 2002, \mn@doi [\aap] {10.1051/0004-6361:20020682},
  \href {https://ui.adsabs.harvard.edu/abs/2002A&A...389..931T} {389, 931}

\bibitem[\protect\citeauthoryear{{\VAN{Van}{Van}{van}}~Genderen}{{\VAN{Van}{Van}{van}}~Genderen}{2001}]{V01}
{\VAN{Van}{Van}{van}}~Genderen A.~M.,  2001, \mn@doi [\aap]
  {10.1051/0004-6361:20000022}, \href
  {https://ui.adsabs.harvard.edu/abs/2001A&A...366..508V} {366, 508}

\bibitem[\protect\citeauthoryear{{\VAN{Van}{Van}{van}}~Genderen, {Sterken}  \&
  {de Groot}}{{\VAN{Van}{Van}{van}}~Genderen et~al.}{1998}]{V98}
{\VAN{Van}{Van}{van}}~Genderen A.~M.,  {Sterken} C.,   {de Groot} M.,  1998,
  \aap, \href {https://ui.adsabs.harvard.edu/abs/1998A&A...337..393V} {337,
  393}

\bibitem[\protect\citeauthoryear{Walborn, Gamen, Morrell, Barb{\'{a}},
  Laj{\'{u}}s  \& Angeloni}{Walborn et~al.}{2017}]{W17}
Walborn N.~R.,  Gamen R.~C.,  Morrell N.~I.,  Barb{\'{a}} R.~H.,  Laj{\'{u}}s
  E.~F.,   Angeloni R.,  2017, \mn@doi [The Astronomical Journal]
  {10.3847/1538-3881/aa6195}, 154, 15

\bibitem[\protect\citeauthoryear{{Weis}, {Duschl}  \& {Bomans}}{{Weis}
  et~al.}{2003}]{W03}
{Weis} K.,  {Duschl} W.~J.,   {Bomans} D.~J.,  2003, \mn@doi [\aap]
  {10.1051/0004-6361:20021680}, \href
  {https://ui.adsabs.harvard.edu/abs/2003A&A...398.1041W} {398, 1041}

\makeatother
\end{thebibliography}
\appendix
\section{Emission lines used to estimate the radial velocity of LBVs}
\label{sec:appendix}
Below we show all the emission lines that we used in this work to estimate the radial velocity of LBVs. Each figure shows all epochs corresponding to a target. The earliest spectrum is shown in pink and the spectrum evolves over time from the bottom to the top of the figure (pink to brown). Estimated heliocentric emission velocity corresponding to each epoch is marked in a dashed colored line matching the color of the spectrum. In most cases, it is hard to see the individual emission velocities because they overlap with the average heliocentric emission velocity which is shown
in a solid black line (listed in Table~\ref{tab:FinalVelocity}). We should emphasize that in all cases the emission velocity for each target does not change significantly over time.

The local rotation curve velocities are shown in dashed grey lines. In some cases like LBV R110,  emission and rotation curve velocities overlap. However, note that in most cases, 
the difference between average emission velocity and the rotation curve velocity ($\Delta \rm{v_r}$, listed in Table~\ref{tab:FinalVelocity}) is much larger than the differences in emission velocity we measure. For example, Fig.~\ref{fig:sdorall} shows spectra of LBV S~Dor over time. Even though the profile changes slightly from a P Cygni profile to a very minor inverse P Cygni profile, the emission velocities are consistent, and $\Delta \rm{v_r}$ is much larger than the slight change in the emission velocity estimates.

\begin{figure}
\subfloat[]{
  \includegraphics[width=0.45\textwidth]{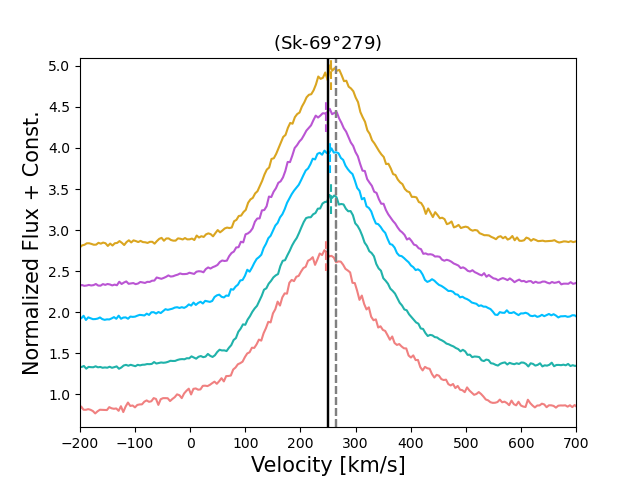}}
\newline
\subfloat[]{
  \includegraphics[width=0.45\textwidth]{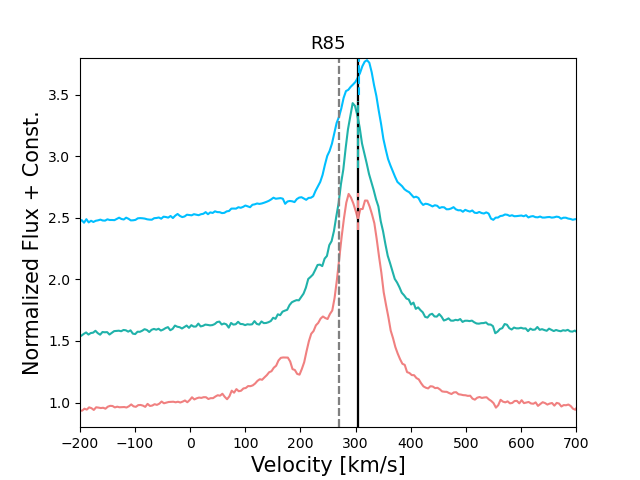}}
\caption{H$\alpha$\ emission lines used to estimate the radial velocity of LBVs. Each figure shows multi-epoch spectra of a target evolving over time from bottom to top. The heliocentric emission velocity and local rotation curve velocity are shown in solid black and dashed grey lines respectively.}\label{fig:Balmeremission}
\end{figure}

\begin{figure}
\subfloat[]{
  \includegraphics[width=0.45\textwidth]{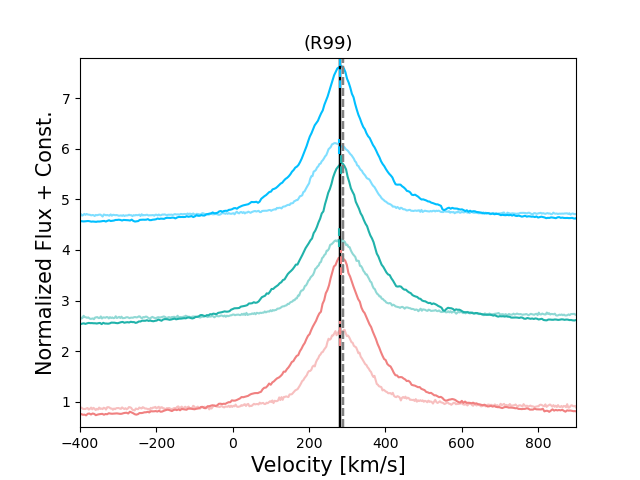}}
\newline
\subfloat[]{
  \includegraphics[width=0.45\textwidth]{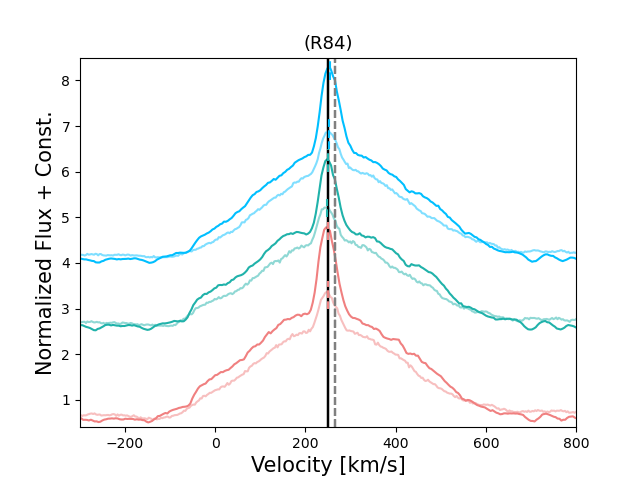}}
\newline
\subfloat[]{
  \includegraphics[width=0.45\textwidth]{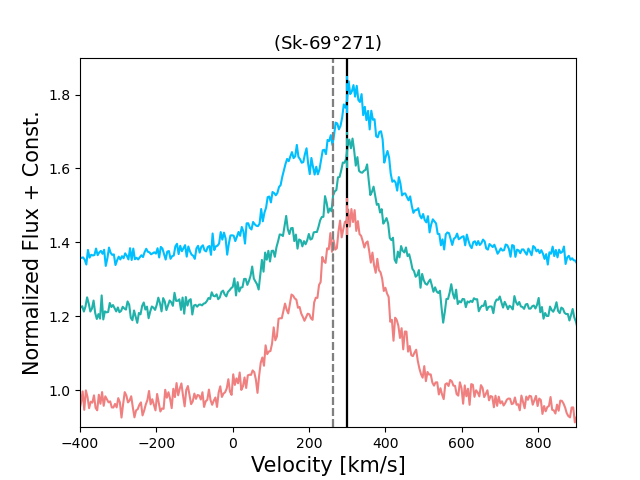}}
\caption{H$\alpha$\ and H$\beta$\ lines are shown in darker and lighter shade lines respectively. For LBV R99, H$\gamma$\ and He~{\sc i} lines are also used but they are not shown here to minimize clutter in the display.}\label{fig:Balmeremission2}
\end{figure}

\begin{figure}
\subfloat[Si~{\sc ii} lines.]{
  \includegraphics[width=0.45\textwidth]{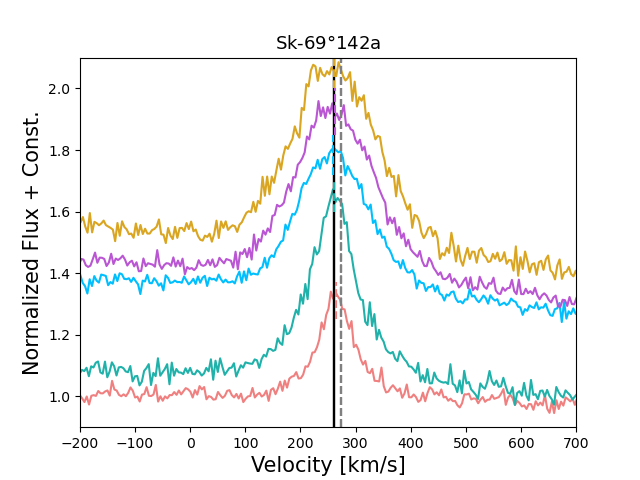}

}
\newline
\subfloat[Si~{\sc ii} lines at $\lambda$5979 \AA \ and $\lambda$5958 \AA \ are shown in darker and lighter shade lines respectively.]{
  \includegraphics[width=0.45\textwidth]{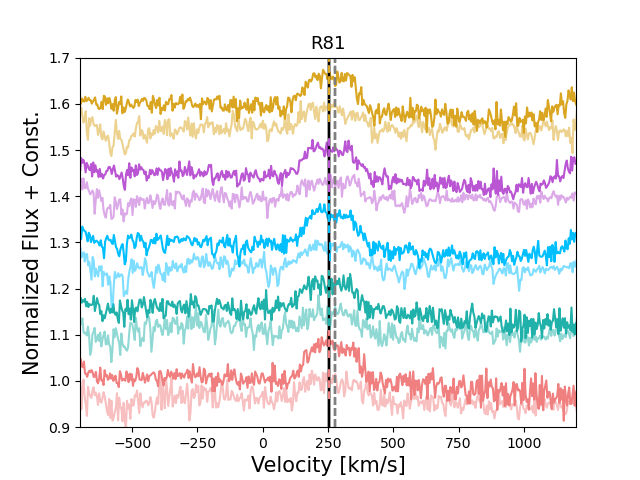}

}
\newline
\subfloat[Si~{\sc iv} lines.]{
  \includegraphics[width=0.45\textwidth]{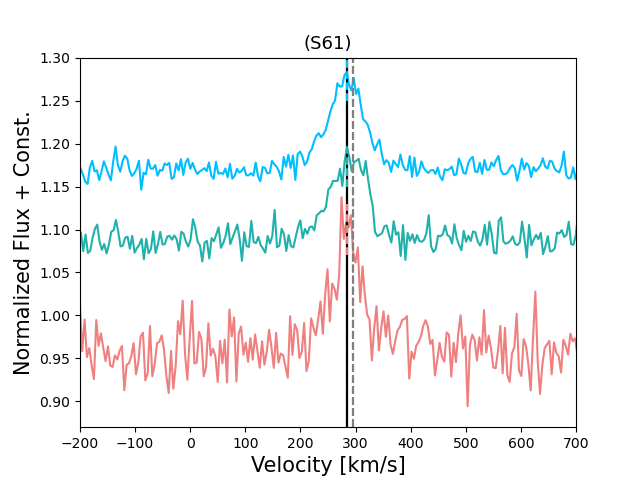}

}
\caption{Same as Fig.~\ref{fig:Balmeremission} but using Si~{\sc ii} or Si~{\sc iv} lines.} \label{fig:Siemission}
\end{figure}

\begin{figure}
\subfloat[He~{\sc ii} lines.]{
  \includegraphics[width=0.45\textwidth]{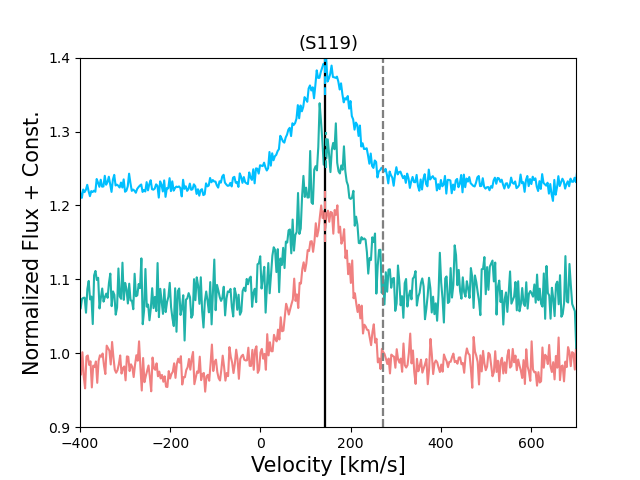}

}
\newline
\subfloat[He~{\sc i} and N~{\sc iii} (lighter) lines.]{
  \includegraphics[width=0.45\textwidth]{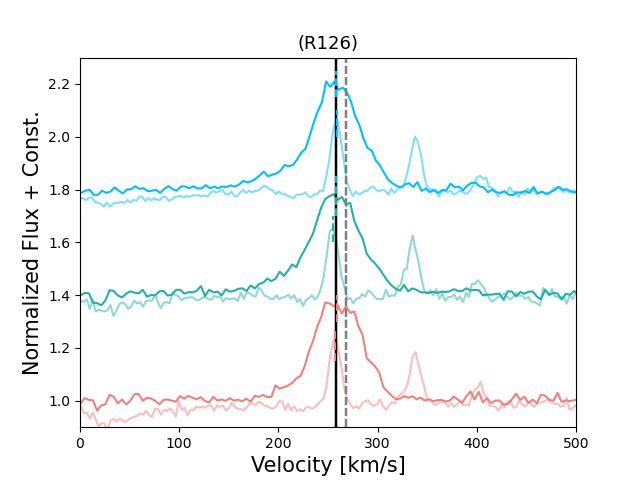}

}
\newline
\subfloat[Mg~{\sc ii} lines at $\lambda$7877 \AA \ and $\lambda$7896 \AA \ (lighter) lines.]{
  \includegraphics[width=0.45\textwidth]{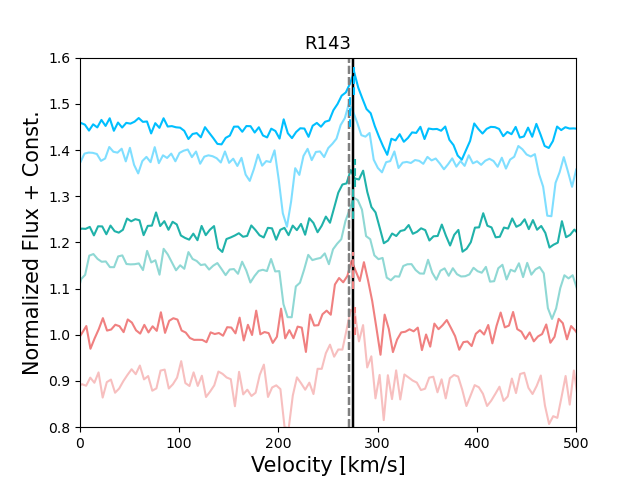}

}
\caption{Same as Fig.~\ref{fig:Balmeremission} but using different lines.}
\end{figure}

\begin{figure}
\subfloat[{[}Fe~{\sc ii}{]} lines.]{
  \includegraphics[width=0.45\textwidth]{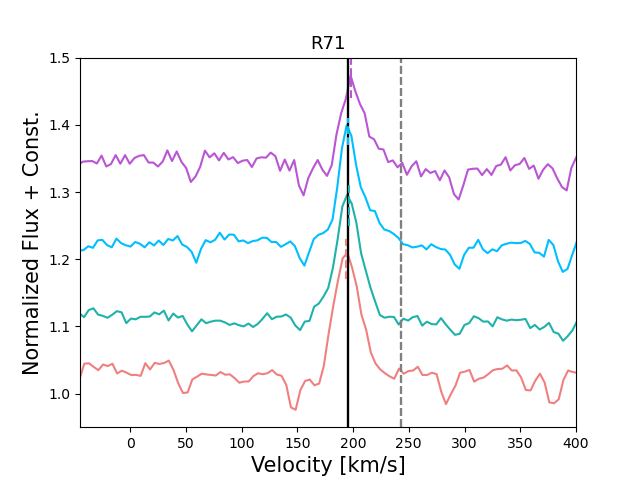}}
\newline
\subfloat[Fe~{\sc ii} and {[}N~{\sc ii}{]} (lighter) lines.]{
  \includegraphics[width=0.45\textwidth]{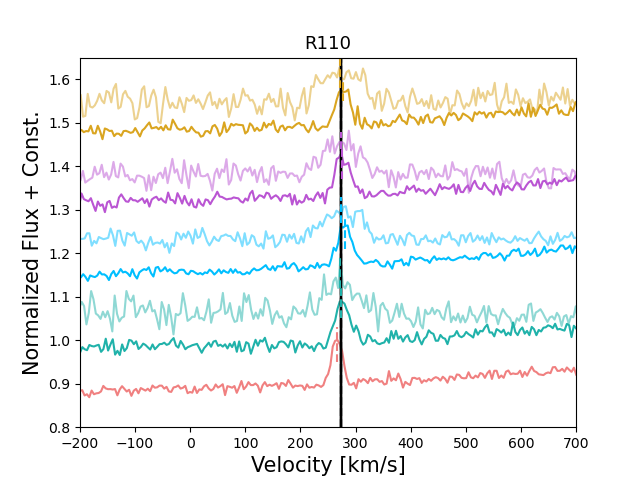}}
\newline
\subfloat[Fe~{\sc ii} and {[}Fe~{\sc ii}{]} (lighter) lines. Some epochs exhibit P Cygni or inverse P Cygni features. To test how much this affects the velocity dispersion of LBVs, we exclude this target and we find that the velocity dispersion of LBVs is consistent with the results shown in Figs.~\ref{fig:LBVRotationCurveEmission} and \ref{fig:LBVcornerRotationCurveEmission}.]{
  \includegraphics[width=0.45\textwidth]{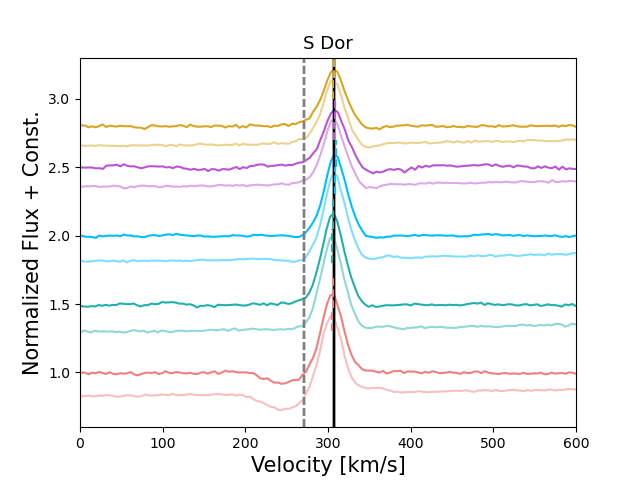}}
\caption{Same as Fig.~\ref{fig:Balmeremission} but using Fe lines.}\label{fig:sdorall}
\end{figure}

\begin{figure}
\centering
\includegraphics[width=0.95\linewidth]{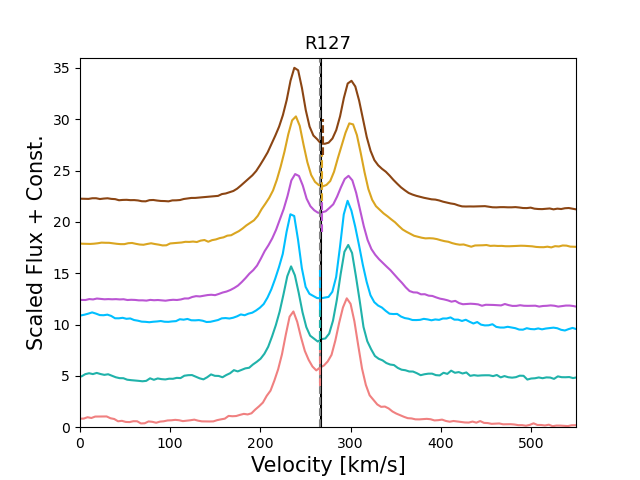}
\caption{Using nebular [N~{\sc ii}] lines. Some of the epochs are scaled by a constant. The relative strength of the peaks varies over time because spectra are extracted at different positions in the nebula due to different slit position angle, not because the nebula is actually changing over time.}\label{fig:Forbid}
\end{figure}

\label{lastpage}
\end{document}